\documentclass[a4paper,superscriptaddress,twocolumn,pre,showpacs,amsmath,amssymb]{revtex4}

\usepackage[T1]{fontenc}
\usepackage[latin1]{inputenc}
\usepackage{color}
\usepackage{vmargin}
\usepackage{graphicx}
\usepackage{txfonts}

\setmarginsrb{20mm}{20mm}{20mm}{15mm}{15pt}{11mm}{0pt}{11mm}%

\renewcommand{\vec}[1]{\boldsymbol{#1}}
\newcommand{\grad}[1]{\vec{\nabla}{#1}}
\newcommand{\curl}[1]{\vec{\nabla}\times{#1}}
\renewcommand{\div}[1]{\vec{\nabla}\cdot{#1}}

\newcommand{\f}[2]{\frac{#1}{#2}}
\newcommand{\dpart}[2]{\frac{\partial #1}{\partial #2}}

\newcommand{\rey}{\mathrm{Re}}
\newcommand{\reym}{\mathrm{Rm}}

\begin{document}
%\nocite{*}
\title{Periodic magnetorotational dynamo action as a prototype
of\\ nonlinear magnetic field generation in shear flows}

\author{J. Herault}
\affiliation{Universit\'e de Toulouse; UPS-OMP; IRAP; Toulouse, France}
\affiliation{CNRS; IRAP; 14, avenue Edouard Belin, F-31400 Toulouse, France}
\affiliation{Laboratoire de Physique Statistique de l'Ecole Normale Sup\'erieure, CNRS UMR 8550,
24 Rue Lhomond, 75231 Paris Cedex 05, France}
\author{F. Rincon}
\email{rincon@ast.obs-mip.fr}
\affiliation{Universit\'e de Toulouse; UPS-OMP; IRAP; Toulouse, France}
\affiliation{CNRS; IRAP; 14, avenue Edouard Belin, F-31400 Toulouse, France}
\author{C. Cossu}
\affiliation{CNRS-Institut de M\'ecanique des Fluides de Toulouse (IMFT), All\'ee du Professeur Camille Soula, 31400 Toulouse, France}
\author{G. Lesur}
\affiliation{UJF-Grenoble 1 / CNRS-INSU, Institut de Plan\'etologie et d'Astrophysique de Grenoble (IPAG) UMR 5274, Grenoble, F-38041, France}
\affiliation{Department of Applied Mathematics and Theoretical Physics,
University of Cambridge,\\ Centre for Mathematical Sciences, Wilberforce
Road, Cambridge CB3 0WA, United Kingdom}

\author{G. I. Ogilvie}
\affiliation{Department of Applied Mathematics and Theoretical Physics,
University of Cambridge,\\ Centre for Mathematical Sciences, Wilberforce
Road, Cambridge CB3 0WA, United Kingdom}
\author{P.-Y. Longaretti}
\affiliation{UJF-Grenoble 1 / CNRS-INSU, Institut de Plan\'etologie et d'Astrophysique de Grenoble (IPAG) UMR 5274, Grenoble, F-38041, France}

\date{\today}

\begin{abstract}
The nature of dynamo action in shear flows prone to
magnetohydrodynamic instabilities is investigated using the
magnetorotational dynamo in Keplerian shear flow as a prototype
problem. Using direct numerical simulations and Newton's method, we
compute an exact time-periodic magnetorotational dynamo solution to
the three-dimensional dissipative incompressible magnetohydrodynamic
equations with rotation and shear. We discuss the physical mechanism
behind the cycle and show that it results from a combination of
linear and nonlinear interactions between a large-scale axisymmetric
toroidal magnetic field and non-axisymmetric perturbations amplified
by the magnetorotational instability.
We demonstrate that this large-scale dynamo mechanism is
overall intrinsically nonlinear and not reducible to the standard
mean-field dynamo formalism. Our results therefore provide clear evidence
for a generic nonlinear generation mechanism of time-dependent
coherent large-scale magnetic fields in shear flows and call for new
theoretical dynamo models. These findings may offer important clues to
understand the transitional and statistical properties of subcritical
magnetorotational turbulence.
\end{abstract}


\maketitle

\section{Introduction}
The generation of coherent, system-scale magnetic fields
in flows of electrically conducting fluids is a long-standing
magnetohydrodynamics (MHD) problem which has so far mostly been
analyzed in terms of linear, kinematic mean-field dynamo
action \citep{moffatt77}. Mean-field effects have in particular long
been invoked to explain the origin of magnetic cycles in MHD rotating
shear flows \citep{parker55,branden05}. 
However, there is currently no
mathematical theory and only little physical understanding of
mean-field dynamos in parameter regimes (kinetic and magnetic
Reynolds numbers) typical of laboratory dynamo experiments and natural
MHD shear flows. Overall, the physical nature of dynamo action in
such flows remains a mostly open question, with critical
implications for geophysics and astrophysics.

Studying the dynamics of shear flows prone to the development
of various local three-dimensional (3D) MHD instabilities is
one of the most promising avenues of research on this problem. 
Numerical simulations of this class of flows have explicitly
demonstrated their potential for coherent dynamo action
\citep{branden95,hawley96,drecker2000,spruit02,cline03,braithwaite06,rincon07b,lesur08,davis10,tobias11},
and it has occasionally been pointed out that their dynamics differs
markedly from that of kinematic mean-field dynamos, as magnetic field
amplification does not proceed exponentially in time and requires
finite-amplitude magnetic perturbations coupling dynamically to fluid
motions. Identifying the physical mechanisms underlying this behaviour
is of prime importance to improve our understanding of dynamo action.

It has recently been realized that these instability-driven
\citep{spruit02,cline03} or subcriti\-cal dy\-na\-mos in shear
  flows \citep{rincon07b,rincon08,rempel10} 
bear many similarities to the hydrodynamic transition to turbulence of shear
flows \citep{hamilton95,waleffe97,faisst04,hof06,eckhardt07}, 
 a fundamentally nonlinear process whose dynamics involves
a variety of hydrodynamic coherent structures such as 
equilibria, travelling waves
\citep{waleffe98,faisst03,wedin04,gibson09}
 or limit cycles
\citep{kawahara01,toh03,viswanath07,halcrow09}. 
This naturally raises the question of the existence and dynamical
relevance of coherent structures in MHD shear flows. Most importantly
for dynamo theory, could time-dependent large-scale dynamo action in
such flows be explained in terms of 3D nonlinear dynamo cycles ? Besides,
if nonlinear dynamo cycles exist for this kind of systems, what can be
learned from them regarding how turbulence originates and develops in MHD
shear flows ?

The problem of magnetorotational (MRI) dynamo action in Keplerian flow
(often referred to as ``zero net flux MRI''), encountered in the context
of astrophysical accretion disks \cite{balbus98}, is particularly
interesting to address these questions, as
direct numerical simulations display pseudo-cyclic
magnetic dynamics \citep{branden95,lesur08,davis10} 
and indicate that transition to sustained MHD turbulence is intrinsically
3D and nonlinear \cite{hawley96}. Besides, numerical MRI dynamo
turbulence  seems to have finite lifetime and to be structured around
a chaotic saddle \citep{rempel10}, much like hydrodynamic turbulence
in pipe flow \citep{faisst04,hof06}. The search for coherent MRI
dynamo structures such as nonlinear cycles is still in its infancy,
though. Progress on these matters is not only desirable from a general
dynamo perspective, it would also shed light on the transitional
properties of MRI turbulence \citep{fromang07b} and on its dynamical
properties (saturation, transport) in fully developed regimes. 

The only exact MRI dynamo structure known as yet is a nonlinear
equilibrium in Keplerian MHD plane Couette flow with walls
\citep{rincon07b}, and it only exists for a limited range of 
parameters. In this work, we report the first accurate calculation of
a 3D nonlinear MRI dynamo limit cycle at moderate (transitional)
kinetic and magnetic Reynolds numbers using numerical techniques
similar to those used for computing nonlinear hydrodynamic cycles in
plane Couette flow \citep{viswanath07}. The study of this nonlinear
MRI dynamo cycle enables us to investigate in detail the essential
physical mechanisms underlying sustained time-dependent MRI
dynamo action. In particular, we show that the cycle
dynamics is not amenable to a standard mean-field dynamo description,
thereby providing clear and detailed evidence for a fully nonlinear
mechanism of coherent magnetic field generation in shear flows prone
to MHD instabilities. 

The paper is organized as follows. In Sect.~\ref{framework}, we
introduce the theoretical framework and numerical methods used in the
study. In Sect.~\ref{excite}, we discuss the physical principles of the MRI
dynamo and describe our strategy to excite large-scale, coherent
recurrent dynamics in direct numerical
simulations. Section~\ref{cycle} is devoted to the presentation
and detailed analysis of the main result of the paper, namely the
discovery of a nonlinear MRI dynamo cycle computed thanks to a Newton
algorithm. A discussion (Sect.~\ref{discussion}) of the results and of
their possible implications for dynamo theory, astrophysics and
research on hydrodynamic shear flows concludes the paper.

\section{Theoretical and numerical framework\label{framework}}
\subsection{The shearing sheet}
\begin{figure}[t]
\resizebox{\hsize}{!}{\includegraphics{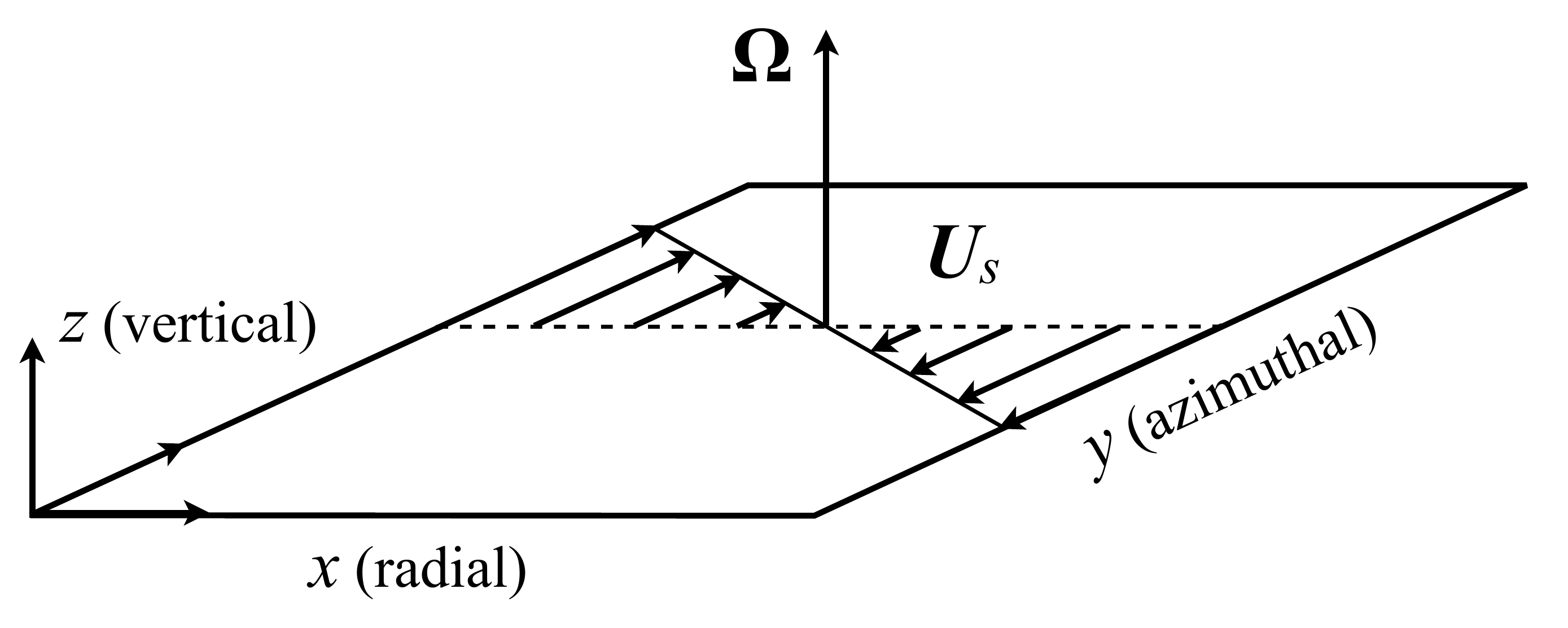}}
  \caption{Geometry of the shearing sheet. For a Keplerian flow, the
    vorticity of $\vec{U}_s$ is anti-aligned with the rotation vector.}
\label{figure1}
\end{figure}

We use the shearing sheet approach to differentially rotating flows
\citep{goldreich65}, whereby a cylindrically symmetric differential
rotation profile is approximated locally by a linear shear flow
$\vec{U}_s=-Sx\,\vec{e}_y$ and a uniform rotation rate
$\vec{\Omega}=\Omega\, \vec{e}_z$. For a Keplerian flow,
$\Omega=(2/3)\, S$. Here, $(x,y,z)$ are respectively the
shearwise, streamwise and spanwise directions (radial, azimuthal 
and vertical in accretion disks). The geometry of the shearing sheet
is represented in Fig.~\ref{figure1}.
To comply with dynamo terminology, we refer to the $(x,z)$ projection
of vector fields as their poloidal component and to their $y$
projection as their toroidal component. ``Axisymmetric'' fields have
no $y$ dependence. For readers familiar with hydrodynamic plane
Couette flow, non-axisymmetric perturbations in our problem
correspond to ``streamwise-dependent'' perturbations in
plane Couette flow.

We consider incompressible velocity perturbations $\vec{u}$ and
magnetic field $\vec{B}$ whose evolutions obey the 3D
dissipative MHD equations in an unstratified shearing sheet:
\begin{eqnarray}
\label{eq:NS}
\dpart{\,\vec{u}}{t}-Sx\,\dpart{\,\vec{u}}{y}+\vec{u}\cdot\grad{\,\vec{u}} &
=& -2\,\Omega\times \vec{u} +S\,u_x\,\vec{e}_y -\grad{\,\Pi}
  \nonumber \\ & &
+\vec{B}\cdot\grad{\vec{B}}+\nu\Delta\,\vec{u}\,,
\end{eqnarray}
\begin{equation}
\label{eq:induc}
\dpart{\vec{B}}{t}-Sx\,\dpart{\vec{B}}{y}=-SB_x\,\vec{e}_y+\curl{\left(\vec{u}\times\vec{B}\right)}+\eta\Delta\vec{B}\,,
\end{equation}
\begin{equation}
\label{eq:div}
\div{\vec{u}}=0\,,\,\,\,\,\div{\vec{B}}=0\,.
\end{equation}
The kinetic and magnetic Reynolds numbers are defined according to
$\rey=SL^2/\nu$ and $\reym=SL^2/\eta$, where $\nu$ and $\eta$ are
the constant kinematic viscosity and magnetic diffusivity and $L$ is
a typical scale of the spatial domain considered.  $\Pi$ is the total
pressure (including magnetic pressure) and $\vec{B}$ is expressed as an
equivalent Alfv\'en velocity. In the following, velocity and magnetic field 
amplitudes are measured with respect to $S L$, while times are measured 
with respect to the inverse of the shearing  rate $S^{-1}$. 

\subsection{Strategy for capturing cycles\label{frameworkstrategy}}

Eqs.~(\ref{eq:NS})-(\ref{eq:div}) are often implemented numerically
using the cartesian shearing box model described in
Sect.~\ref{frameworkSB} below. In the regime of Keplerian differential
rotation, MRI dynamo action has been found in a large number of
independent direct numerical simulations of this kind
\citep{branden95,hawley96,fleming00,fromang07b,lesur08}, sometimes
in the form of pseudo-cyclic dynamics.
Our strategy to capture a nonlinear MRI dynamo cycle in this regime
follows that used in Ref.~\cite{viswanath07} to compute nonlinear
cycles in hydrodynamic plane Couette flow. First, we tried to excite
pseudo-cyclic dynamics in direct shearing box numerical simulations
(DNS) by devising initial conditions inspired by our partial
knowledge of the underlying physics. Once this was achieved, we
attempted to converge to a cycle using a Newton solver seeded with a
well-chosen DNS snapshot and an estimate for the cycle period. The
first part of this program is described in
Sect.~\ref{excite}. Convergence to a cycle using Newton's method
is presented in Sect.~\ref{cycle}. The numerical methods for achieving
this result are explained below.

\subsection{Numerical method for direct numerical
  simulations\label{frameworkSB}}
The numerics presented in this paper are done in the incompressible 
cartesian shearing box framework, which assumes simple spatial
periodicity in the $y$ and $z$ directions and shear-periodicity in
the $x$ direction. The latter amounts to assuming that a linear
shear flow is constantly imposed and that all quantities are periodic
in a sheared Lagragian frame. This approximation is justified in the
context of centrifugally supported differentially rotating flows, such
as astrophysical accretion flows. Simulations of this kind are also
sometimes referred to as homogeneous shear flow turbulence simulations
(see e.g. Ref.~\citep{pumir96,gualtieri02}). 

Time integrations (direct numerical simulations) are carried
out in a shearing box of size $(L_x,L_y,L_z)$ using the SNOOPY code
\citep{lesur07}. The numerical time integration scheme used 
is a standard explicit third-order Runge-Kutta algorithm. The code
relies on a spectral implementation of the shearing box model similar
to that described in Ref.~\cite{umurhan04}. A discrete spectral basis
of ``shearing waves'' (or Orr-Kelvin waves, see
Refs.~\cite{kelvin1887,orr1907}) with constant $k_y$ and $k_z$
wavenumbers and constant shearwise Lagrangian wavenumber $k_x^{(0)}$
is used to represent the various fields in the sheared Lagragian
frame. The shearing of non-axisymmetric perturbations in this model
is described using time-dependent Eulerian shearwise wavenumbers, 
\begin{equation}
\label{eq:shwave}
k_x(t)=k_x^{(0)}+k_ySt~.
\end{equation}
This equation for the radial wavenumber provides an exact description
of the evolution of non-axisymmetric waves with initially leading
polarization ($k_x^{(0)}\,k_y < 0$) into trailing waves
($k_x(t)\,k_y > 0$, corresponding to a trailing spiral in
cylindrical geometry) under the action of shear. 

An important comment is in order at this stage. If we were to simply
consider the evolution of a given initial set of such waves, we would
only be able to observe dynamics that decay on long times. Indeed,
the fate of all shearing waves is to evolve into strongly trailing
structures with ever smaller scale in $x$, and such structures are
extremely efficiently dissipated by viscous and resistive processes (see
e.g. Ref.~\citep{knobloch85,korycansky92}). In order to accomodate for
possible physical interactions leading to long-lived nonlinear
dynamics in numerical shearing box simulations, a procedure must
therefore be used that leaves open the possibility of physical
generation and dynamical evolution of new leading non-axisymmetric
structures in the course of the simulations. The solution to this
problem is to regularly ``remap'' the basis of shearing waves used to
describe the various fields. At regular time intervals during the
simulations, the energy content of strongly trailing shearing waves is
set to zero, the corresponding basis vector is pruned and replaced by
a new shearing wave basis vector with strongly leading wavenumber 
(see Ref.~\cite{lesurphd}, Chap. 5, Sect.~4). If the simulation is 
well resolved spatially (as is the case for the
results presented in this manuscript), the energy contained into
strongly trailing waves when they are pruned should be negligible (as
a result of their enhanced dissipation), and the remap procedure
should not therefore artificially affect the dynamical evolution of
the system in any significant way. A way to check this is to compare
the energy lost by this procedure with the  energy dissipated by
viscous/resistive diffusion. We always find that artificial energy
losses are negligible in SNOOPY for spatially well-resolved
simulations \citep{lesur05}. We also point out that the remap
procedure does not by itself inject energy into new leading waves but
merely provides room for them in the wavenumber grid. Finally, we
emphasize that all nonlinearities of the shearing sheet MHD
Eqs.~(\ref{eq:NS})-(\ref{eq:div}), including nonlinear interactions
between all the shearing waves present at a given spatial resolution,
are retained in our numerical model. A standard pseudo-spectral
method with dealiasing is used to compute all nonlinear terms at 
each time step. 

\subsection{Newton's method for computing nonlinear cycles}
Newton's method is a standard tool for computing nonlinear coherent
structures such as saddle points, travelling waves, or nonlinear
cycles in high-dimensional dynamical systems. In recent years, the
method has been applied successfully to the three-dimensional
Navier-Stokes equations for various wall-bounded shear flows
\citep{waleffe98,faisst03,wedin04,viswanath07,gibson09,halcrow09}
and to the MHD equations in Keplerian plane Couette flow
\cite{rincon07b}.
For the purpose of this study, we developed a new Newton solver
called PEANUTS. The solver makes use of the PETSc toolkit \citep{petsc}
and is based on an efficient matrix-free Newton-Krylov algorithm
particularly well adapted to calculations for high-dimensional
dynamical systems such as those resulting from the discretization of
the three-dimensional partial differential equations of fluid dynamics. 
It can be used to compute nonlinear equilibria, travelling waves
and limit cycles for a variety of partial differential
equations. For a nonlinear cycle search, the code minimizes
$||\vec{X}(T)-\vec{X}(0)||_2/||\vec{X}(0)||_2$, where $\vec{X}(t)$ is
a state vector containing all independent
field components at time $t$, and $T$ is a guess
for the period \citep{viswanath07}.  An eigenvalue solver
based on the SLEPc toolkit \citep{slepc} was implemented to compute
the stability of nonlinear states. The code was tested against
solutions to the Kuramoto-Sivashinsky equation \citep{lan08} before
being implemented for the 3D MHD equations in the shearing box, using
SNOOPY as time integrator.

\section{Excitation of recurrent dynamics\label{excite}}
In this Section, we describe our strategy to approach a nonlinear
MRI dynamo cycle using DNS of the 3D MHD equations in the shearing
box. We first discuss in detail what is the ``minimal'' set of initial
conditions required to excite a long-lived MRI dynamo in direct
numerical simulations. We then explain how smooth-enough pseudo-cyclic
dynamics can be excited at moderate Re and Rm with this kind of
initial conditions by varying the aspect ratio of the simulations and
by restricting the dynamics to an invariant subspace associated with a
natural symmetry of the original equations.

\subsection{Devising a good initial guess for a Newton search}
\begin{figure}[t]
\resizebox{\hsize}{!}{\includegraphics{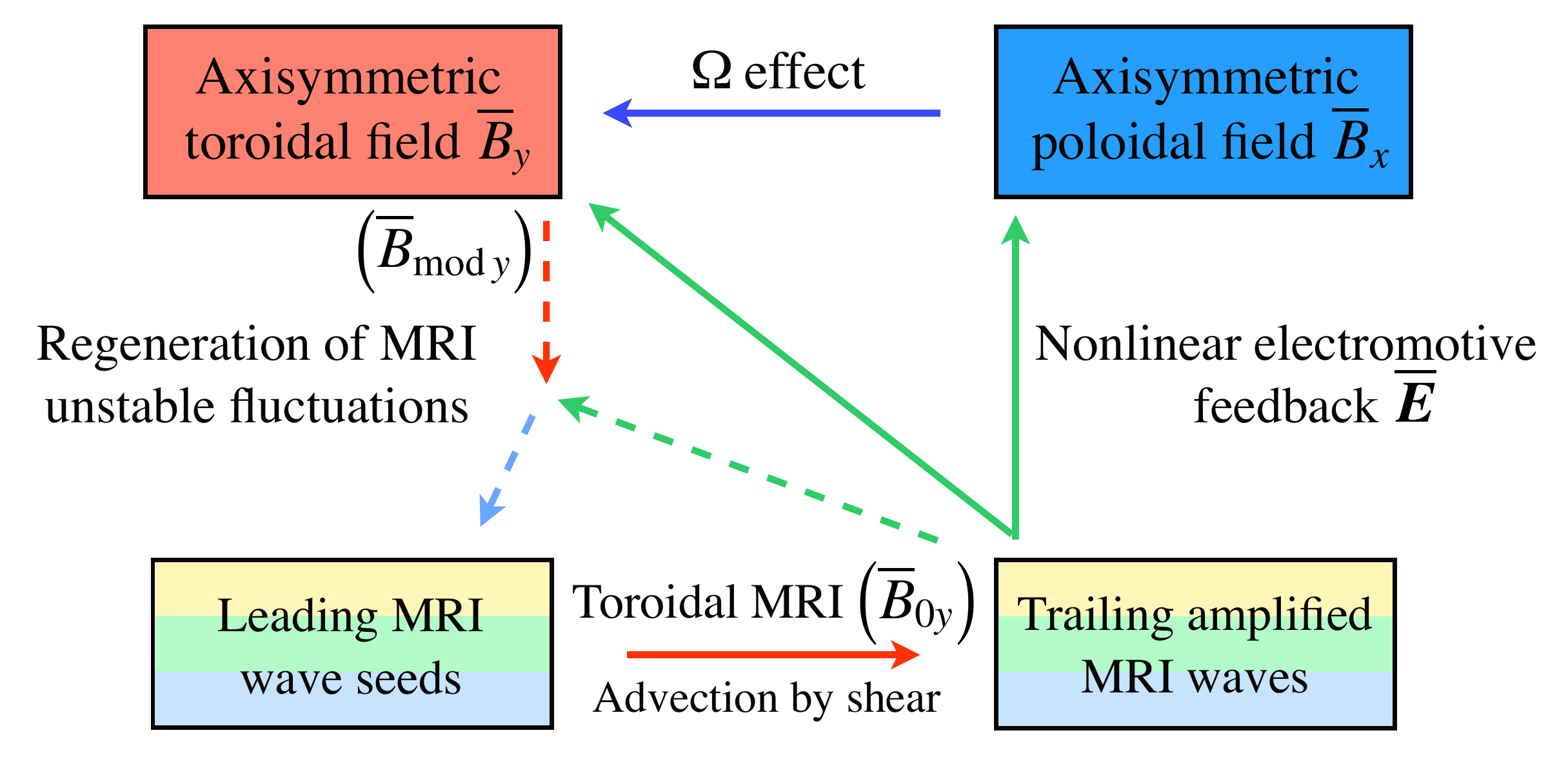}}
  \caption{(Color online) Suggested physical mechanism of the MRI dynamo. 
Full arrows: main dynamo loop. Dashed arrows: nonlinear regeneration of
  MRI-unstable fluctuations. The various colors are used to identify
  the active modes taking part in the cyclic MRI dynamo described in
  Fig.~\ref{figure5}: red (top left box) and blue (top right box) denote 
  axisymmetric field components, different colors in the bottom boxes 
  denote successive non-axisymmetric MRI waves.}
\label{figure2}
\end{figure}

Previous work has demonstrated that instability-driven dynamo action
requires a dynamical interplay between a ``large-scale'' axisymmetric,
instability-supporting magnetic field and  perturbations unstable to
non-axisymmetric MHD instabilities, whose amplification to
nonlinear levels may generate an electromotive force (EMF) with the
ability to sustain the large-scale field
\citep{spruit02,cline03,braithwaite06,rincon07b,rincon08,lesur08,lesur08b,tobias11}. 
The basic processes thought to be responsible for MRI dynamo action
are described below and in Fig.~\ref{figure2}. 
Let us focus on the time-evolution of the axisymmetric magnetic field
$\overline{\vec{B}}(x,z,t)$, where the overbar denotes an average over
$y$, and $\int_0^{L_z}\int_0^{L_x}\overline{\vec{B}}(x,z,t)\,dx\,dz=0$ for
the MRI dynamo (``zero net-flux'') problem. The induction equation for
$\overline{\vec{B}}$ reads
\begin{equation}
  \label{eq:inducaxi}
  \dpart{\overline{\vec{B}}}{t}=-S\,\overline{B}_{x}\,\vec{e}_y+\overline{\curl{\vec{E}}}
+\eta\Delta\overline{\vec{B}}~.
\end{equation}
The first term on the r.h.s. describes the stretching of the axisymmetric
poloidal field into an axisymmetric toroidal field (the so-called
$\Omega$ effect in dynamo theory), the second term is a nonlinear induction
term involving the axisymmetric projection 
$\overline{\vec{E}}=\overline{\vec{u}\times\vec{B}}$ of the
electromotive force resulting from the nonlinear coupling
of velocity and magnetic perturbations, and the third term is the
magnetic diffusion term. In the following, we will be particularly
interested in the time evolution of the fundamental Fourier mode
in $z$  of $\overline{\vec{B}}$, defined as
\begin{equation}
  \label{eq:axifieldfundamental}
\overline{\vec{B}}_0(z,t)=\overline{\vec{B}}_0(t)\cos{\left(k_{z0}\, z\right)}~
\end{equation}
with $k_{z0}=2\pi/L_z$, as this mode is always and by a large amount
the dominant contribution to the total axisymmetric field for the type
of dynamics excited by the class of symmetric initial conditions
described below (a large-scale field with an arbitrary phase in $z$
can of course be excited if the simulation is initialized with
non-symmetric perturbations or if symmetry-breaking instabilities are
allowed to develop during the simulations). Following
Eq.~(\ref{eq:inducaxi}), we also introduce the nonlinear EMF acting on
$\overline{\vec{B}}_0$,
\begin{equation}
\label{eq:EMFfundamental}
\overline{\vec{E}}_0(z,t)=\overline{\vec{E}}_0(t)\sin{\left(k_{z0}\,z\right)}~.
\end{equation}
The time-evolution of $\overline{\vec{B}}_0(t)$ is given by
\begin{equation}
  \label{eq:inducB0}
  \dpart{\overline{\vec{B}}_0}{t}=-S\,\overline{B}_{0x}(t)\,\vec{e}_y+k_{z0}\,\vec{e}_z\times\overline{\vec{E}}_0(t)
-\eta k_{z0}^2\overline{\vec{B}}_0(t)~.
\end{equation}
The physical interpretation of the various terms on the r.h.s. of this
equation is the same as for Eq.~(\ref{eq:inducaxi}).
Our goal to obtain sustained time-dependent MRI dynamo action in
direct numerical simulations was to excite a dynamo loop 
involving the first two terms on the r.h.s. of Eq.~(\ref{eq:inducB0}), as
depicted in Fig.~\ref{figure2}. To do so, we first attempted to start
from a very simple class of initial conditions combining
\begin{enumerate}
\item  a ``large-scale'', axisymmetric
poloidal magnetic field $\overline{B}_{0x}$ (blue box in
Fig.~\ref{figure2}) whose stretching by the shear  (``$\Omega$
effect", full line blue arrow in Fig.~\ref{figure2}) produces an
MRI-unstable axisymmetric toroidal field
$\overline{B}_{0y}$ (red box in Fig.~\ref{figure2}),
\item non-axisymmetric perturbations subject to joint amplification
  (full line red arrow in Fig.~\ref{figure2}) by a transient toroidal
  MRI \cite{balbus92,ogilvie96,terquem96,lesur08b}
of $\overline{B}_{0y}$ and by a kinematic Orr mechanism (swing
amplification of non-axisymmetric waves, see
Refs.~\cite{orr1907,butler92,johnson07}). These perturbations were
chosen in the form of real random shearing wave packets with a full
spectrum of $k_z$ and a single ``horizontal'' wavenumber pair
$(k_x^{(0)}$, $k_y)=\pm (-2\,\pi/L_x,2\,\pi/L_y)$ with leading
($k_x^{(0)}\,k_y < 0$) polarization (different colors in the bottom
left box of Fig.~\ref{figure2} represent successive individual leading
waves). In the course of their evolution, such perturbations may
generate a nonlinear electromotive feedback $\overline{\vec{E}}_0$
with the ability to sustain $\overline{\vec{B}}_0$ (full line green
arrows in Fig.~\ref{figure2}), thereby closing the main dynamo loop. 
\end{enumerate}

It turns out that, independently of the initial amplitudes of each of
these perturbations, this restricted class of initial conditions can only
trigger transient, short-lived dynamics. As explained in
Sect.~\ref{frameworkSB}, non-axisymmetric perturbations
with a single initial horizontal wavenumber ($k_x^{(0)}$, $k_y$) can
only be amplified for a few shearing times before
they get sheared into a strongly trailing ($k_x(t)\,k_y \gg 0$, bottom
right box with multiple colors in Fig.~\ref{figure2}), rapidly decaying structure
\citep{lesur08b}. Besides, their nonlinear self-interaction cannot
give rise to new non-axisymmetric leading waves,  making it
impossible to sustain $\overline{\vec{B}}_0$ against ohmic diffusion
on long times. Hence, long-lived dynamics can only be excited if a
distinct physical mechanism operates that generates new leading,
transiently MRI-unstable perturbations. Exploring this issue, we then
found that much longer-lived dynamics is obtained as soon as
\begin{itemize}
\item[3.] an $x$-dependent axisymmetric modulation
\begin{equation}
\label{eq:bmed}
\overline{\vec{B}}_\mathrm{mod}(x,z,t)=\overline{\vec{B}}_\mathrm{mod}(t)
\cos\left(2\pi x/L_x\right)\cos\left(k_{z0}z\right)
\end{equation}
of $\overline{\vec{B}}$ is initially added on top of
$\overline{\vec{B}}_0$, i.e.
\begin{equation}
  \label{eq:baxitot}
  \overline{\vec{B}}(x,z,t=0)=\overline{\vec{B}}_0(z,t=0)+
  \overline{\vec{B}}_\mathrm{mod}(x,z,t=0)~.
\end{equation}
\end{itemize}
This simple numerical observation indicated that the physical
mechanism by which new leading shearing waves are generated requires
that $\overline{\vec{B}}$ be modulated along the $x$ direction.
The physical explanation for this behaviour is
rather subtle, though : such a modulation of $\overline{\vec{B}}$
confines MRI-unstable perturbations in $x$ and allows for reflections
of non-axisymmetric waves, somehow taking on the role of walls in a
wall-bounded shear flow (non-axisymmetric instabilities in
wall-bounded shear flows take on the form of global standing modes, see
e.g. Ref.~\cite{waleffe97}). In more mathematical terms, the nonlinear
triad  interaction (shown by dashed arrows in Fig.~\ref{figure2}) of 
the ``confining'' (say with $k_x=-2\pi/L_x,k_y=0$) axisymmetric mode
$\overline{\vec{B}}_\mathrm{mod}$ (red dashed arrow) with a trailing
shearing wave (say $k_x(t)=\pi/L_x$, $k_y=2\pi/L_y$, green dashed arrow)
can seed a new leading $(k_x(t)=-\pi/L_x$, $k_y=2\pi/L_y)$ wave (light
blue dashed arrow). This type of mechanism, which 
has long been suspected to be at work in non-rotating hydrodynamic
shear flow turbulence (see for instance the discussion in
Ref.~\citep{farrell93}) and has more recently been invoked 
in the context of hydrodynamic stability of accretion disks
\citep{lithwick07}, is in fact essential to any sustained
non-axisymmetric dynamics in the shearing box. We did check that the
seeding of new leading waves in the simulations
was not related to our implementation of the remap procedure (see
Sect.~\ref{frameworkSB} and Ref.~\cite{umurhan04}), but had a genuine
physical origin. In particular, the fact that long-lived dynamics 
and new leading waves can only be excited in the simulations if a
$\overline{\vec{B}}_\mathrm{mod}$ component is added to the initial
condition demonstrates that our numerical method does not artificially
inject energy into leading waves.

To summarize this paragraph, using various relative combinations of 
axisymmetric and non-axisymmetric initial conditions composed of
$\overline{\vec{B}}_0$, $\overline{\vec{B}}_\mathrm{mod}$ and a random
leading shearing wave packet, we found it possible to obtain
long-lived MRI dynamo action in direct shearing box simulations for
different regimes. Based on these experiments, we claim that the
essence of the driving mechanism of the MRI dynamo can be fully 
explained in terms of the few generic physical mechanisms described 
above and in Figure~\ref{figure2}.  As will be shown below,
this claim is well supported by the targeted numerical
experiment presented in Sect.~\ref{cycle}.

\subsection{Simplifying the dynamics}
\begin{figure}
\resizebox{\hsize}{!}{\includegraphics{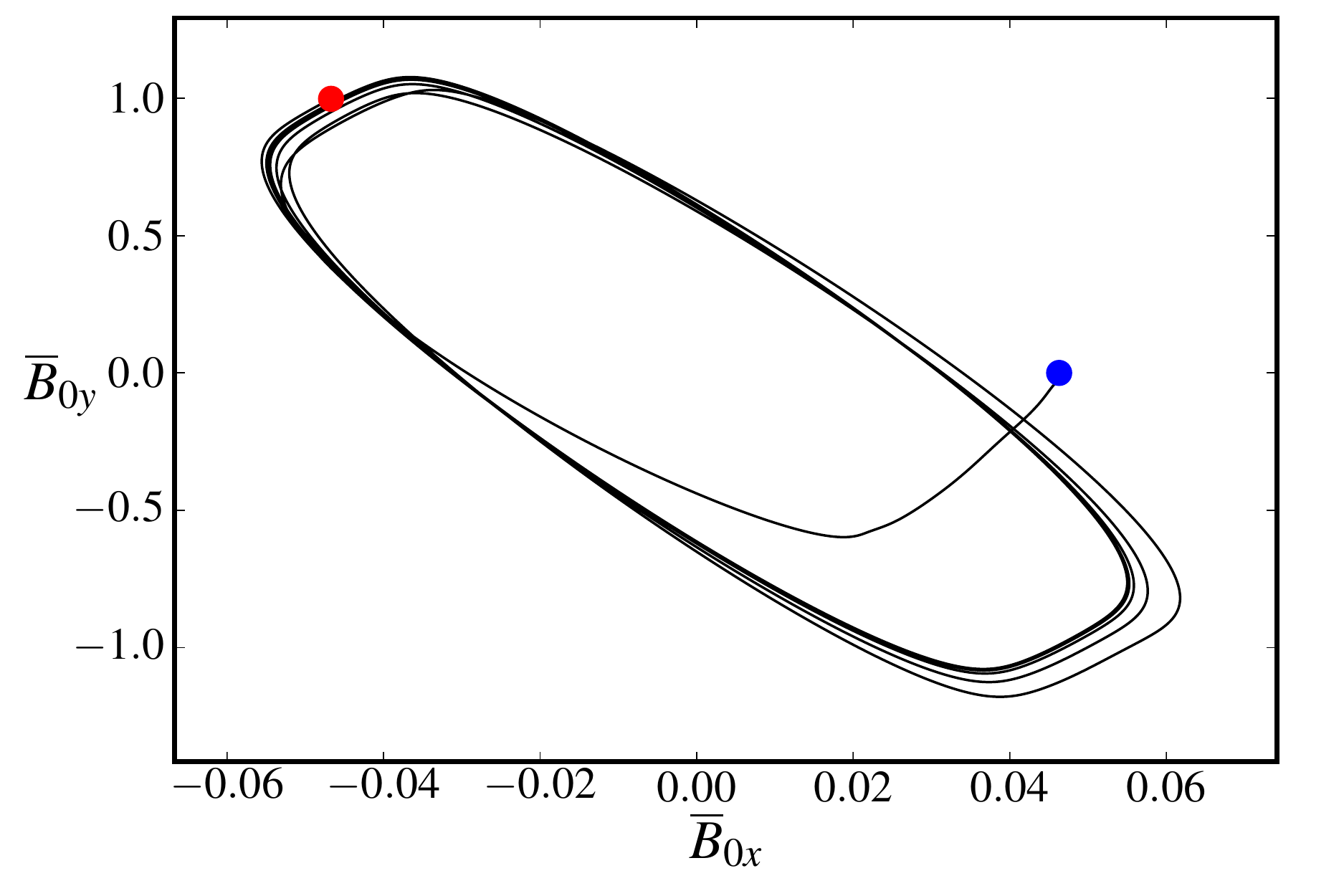}}
\caption{(Color online) Projection in the $\overline{B}_{0x}-\overline{B}_{0y}$
plane of a DNS trajectory approaching a nonlinear cycle
for $(L_x,L_y,L_z)=(0.7,20,2)\,L$ and $\rey=70$, $\reym=360$. 
The full blue circle whose coordinates are $(0.046,0)$ marks the position of the 
system at $t=0$ and the red one located at $(-0.046,0.997)$ marks the position 
at $t=457.1\,S^{-1}$. Typical amplitudes of the various components of the 
initial conditions used to obtain this kind of trajectory are given in 
Sect.~\ref{sensitive}\label{figure3}.}
\end{figure}

Approaching nonlinear cycles by DNS finally required us to find regions of
parameter space in which only a small number of these structures are
present. The dynamics  in regimes ($\rey$, $\reym$ of a few thousands)
typical of simulations displaying pseudo-cyclic dynamics
\citep{branden95,lesur08,davis10} being complex and probably involving
a lot of different coherent structures, we restricted our
investigations to $\rey$, $\reym$ of a few hundreds. In such regimes,
however, shearing waves are quickly damped after they turn trailing,
unless $k_x(t)$ changes on a timescale much longer than
$S^{-1}$. Exciting long-lived dynamics in this $\rey$ and $\reym$ regime
therefore further required setting $L_y\gg L_x$ ($|k_y|\ll |k_x^{(0)}|$ in
Eq.~(\ref{eq:shwave})). Starting from various non-symmetric initial
conditions, we then spotted that nonlinear states approaching a
symmetry $\mathcal{A}_1$ \citep{nagata86} of
the shearing box MHD equations were regularly excited in DNS. 
This symmetry, described in the Appendix, allows for a large-scale
axisymmetric magnetic field with the symmetry of $\overline{\vec{B}}_0$.
In order to isolate these structures more easily, we then enforced
numerically that the dynamics take place in the corresponding
invariant subspace. This strategy was sufficient to excite recurrent
dynamics in a large aspect ratio shearing box with
$(L_x,L_y,L_z)=(0.7,20,2)\,L$ and $\rey=70$, $\reym=360$. A projection
in the $\overline{B}_{0x}-\overline{B}_{0y}$ plane of a DNS trajectory
approaching a nonlinear cycle in this regime is depicted in
Fig.~\ref{figure3}. The trajectory is seen to approach a periodic
orbit after a few tens of shearing times and then to stay
close to it for several hundred shearing times.

\subsection{Sensitive dependence on initial conditions\label{sensitive}}
Before we close this Section, we find it necessary to emphasize that
the dynamical system at hand has a very high sensitivity on initial
conditions. This feature seems to be a very common property of shear flow
turbulence (see for instance Ref.~\cite{faisst04} for the case of
hydrodynamic turbulence in pipe flow) and was explicitly demonstrated
for the MRI dynamo problem in Ref.~\cite{rempel10}. In the course of
this work, we clearly observed that the set of initial conditions
leading to long-lived dynamics for the system at hand is fractal.  
For this reason, any statement of a set of precise numerical values
for the amplitudes of the various types of perturbations involved in
the design of the initial conditions leading to pseudo-cyclic dynamics is 
rather pointless. The problem is that different numerical codes,
initialized exactly in the same way, will almost certainly diverge
after a few shearing times  because their algorithms will generate
different numerical ``noise'' (low amplitude numerical errors) at each
time step, subsequently leading to a rapid divergence
(between different codes and also possibly different computer
architectures) of phase-space trajectories such as that shown in
Fig.~\ref{figure3}. The initial amplitudes and relative mixtures of
modes required to approach nonlinear cycles are therefore in the end
specific to the numerical methods used in the code (this does not imply 
that the cyclic nonlinear solutions are not themselves robust, as 
discussed in the next Section).  The only relevant helpful information 
for a reader eager to reproduce the results presented in Sect.~\ref{cycle} 
and trajectories similar to that shown in Fig.~\ref{figure3} is a set of 
approximate values for the amplitudes of the perturbations entering our 
class of initial conditions, around which we found it possible to excite
recurrent dynamics for $(L_x,L_y,L_z)=(0.7,20,2)\,L$ and $\rey=70$ and
$\reym=360$: $\overline{B}_{0x}\simeq 0.046 SL$,
$\overline{B}_\mathrm{mod\, y}\simeq 0.11 SL$, and non-axisymmetric
shearing wave packets (with random relative amplitudes for different
$k_z$, as described earlier) with  comparable velocity and magnetic
amplitudes of the order $0.3 SL$. 

\begin{figure*}
\resizebox{\hsize}{!}{\includegraphics{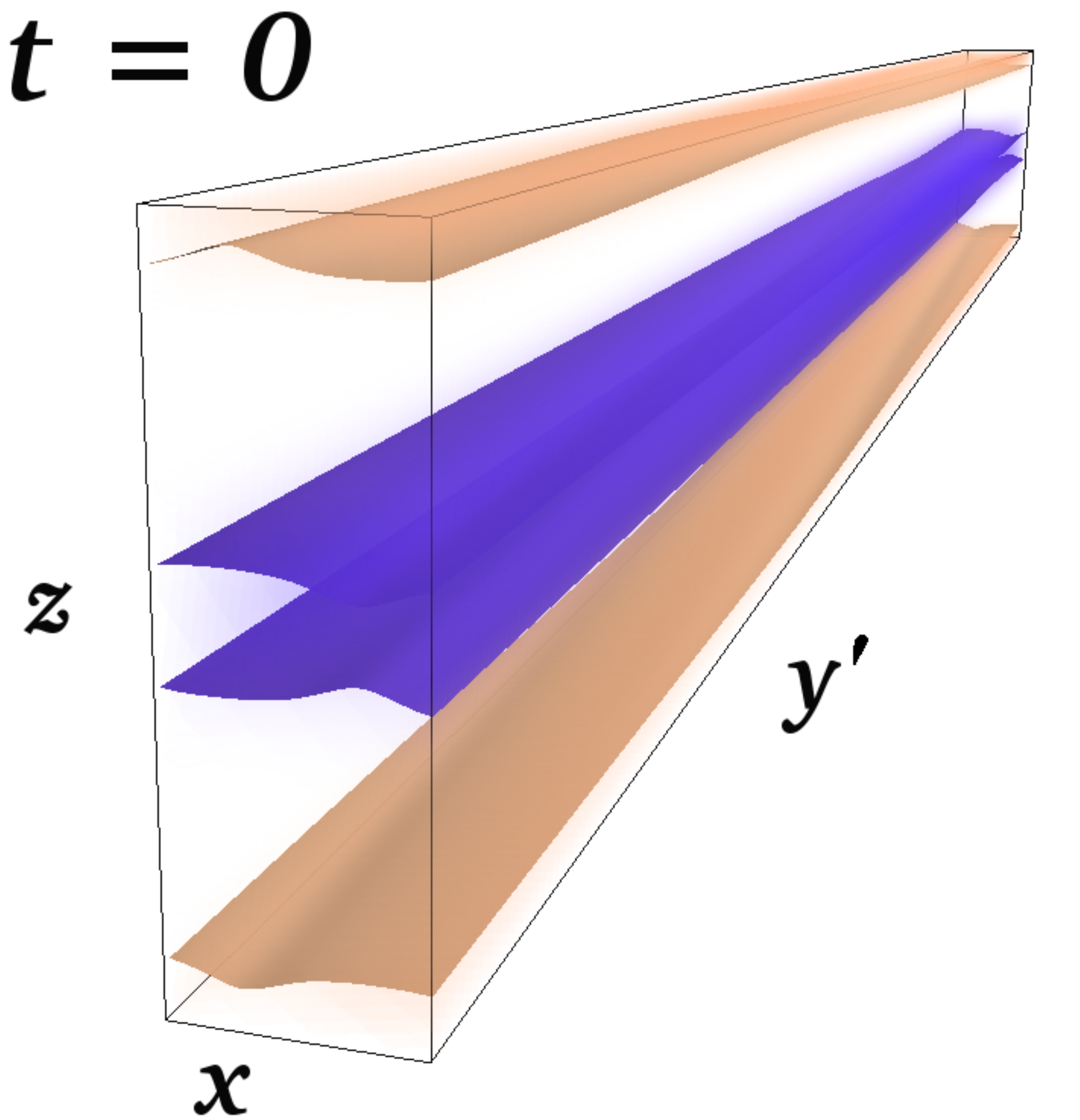}\includegraphics{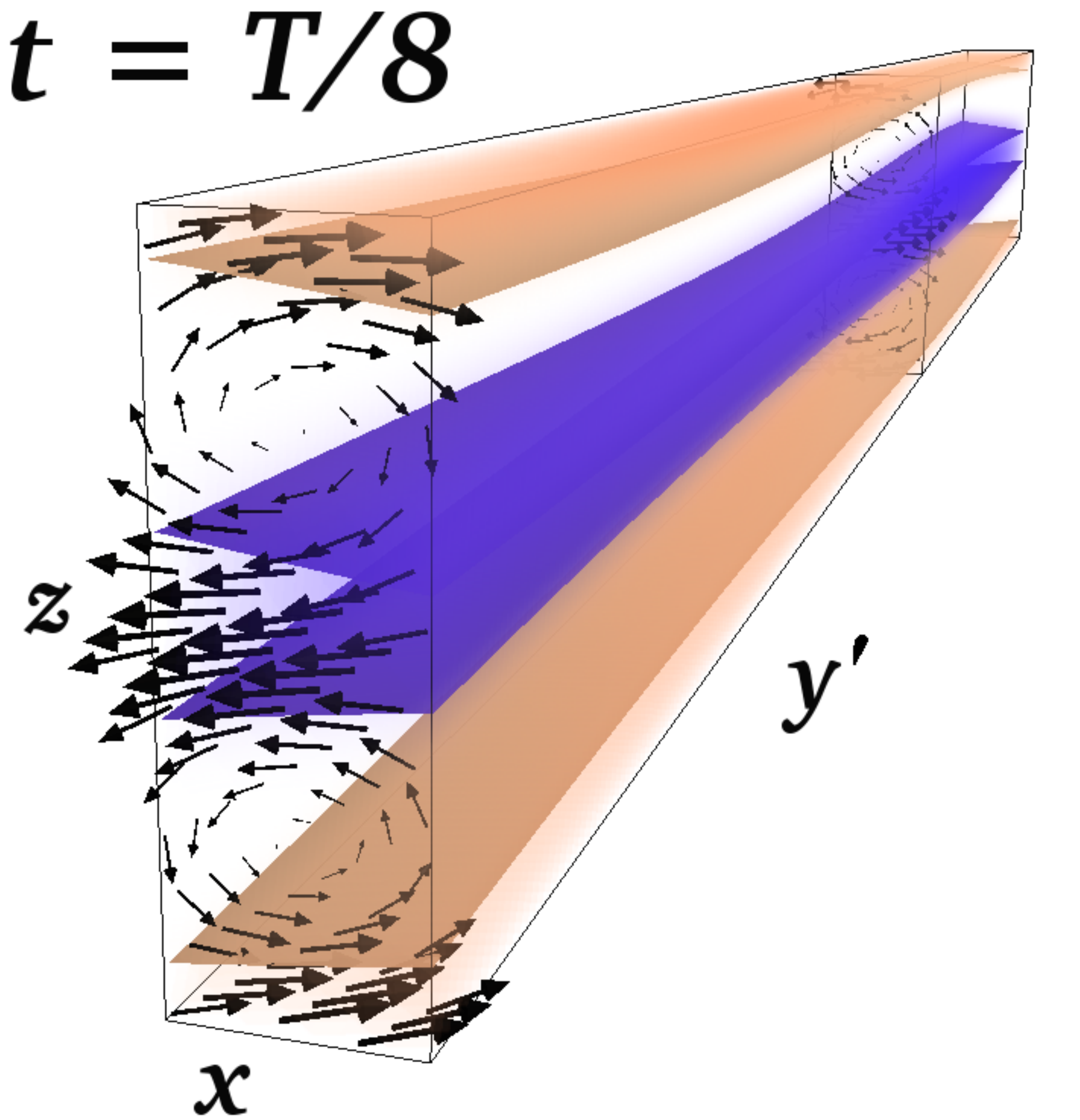}\includegraphics{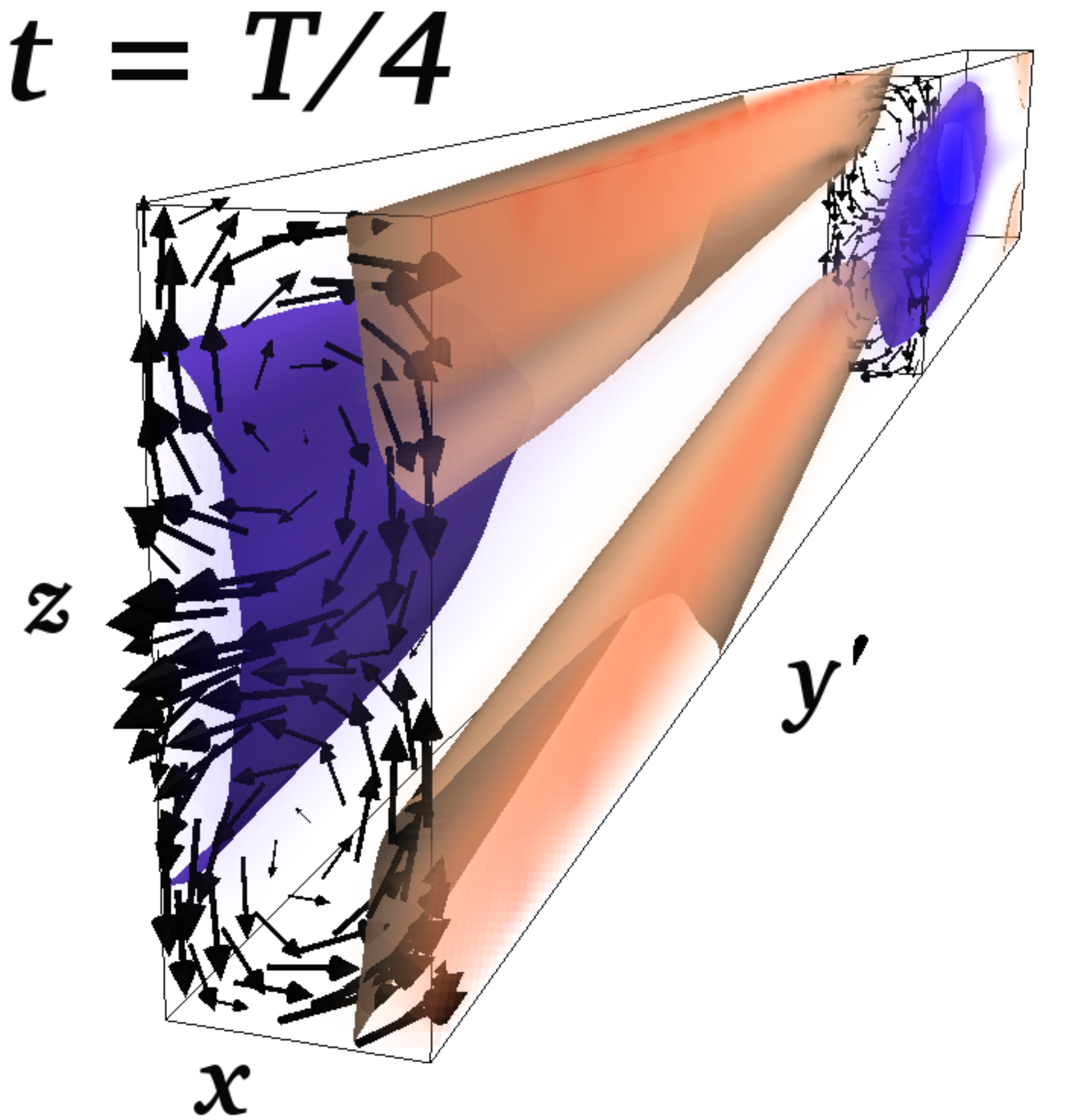}\includegraphics{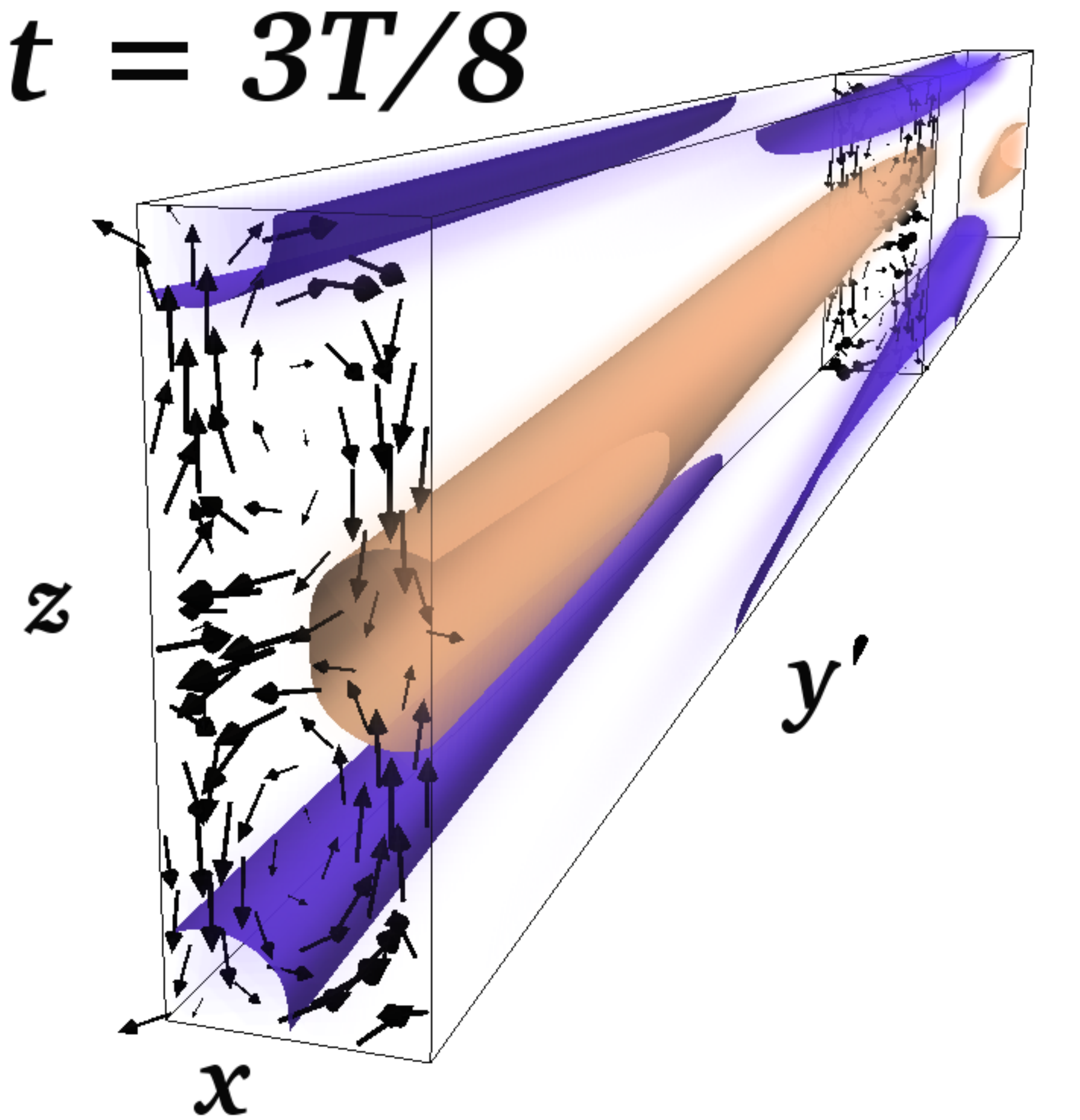}}
\vspace{0.1cm}

\resizebox{\hsize}{!}{\includegraphics{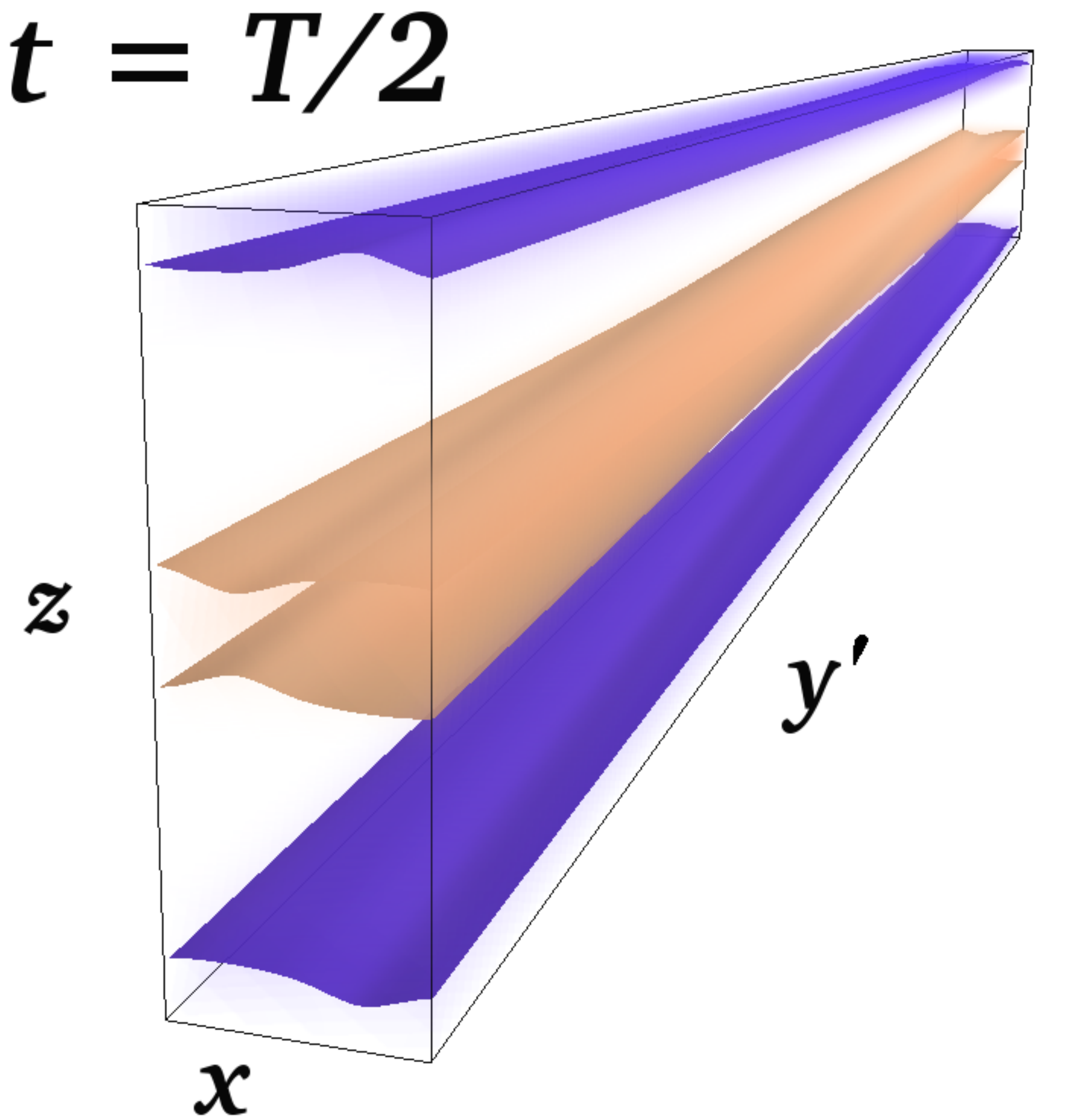}\includegraphics{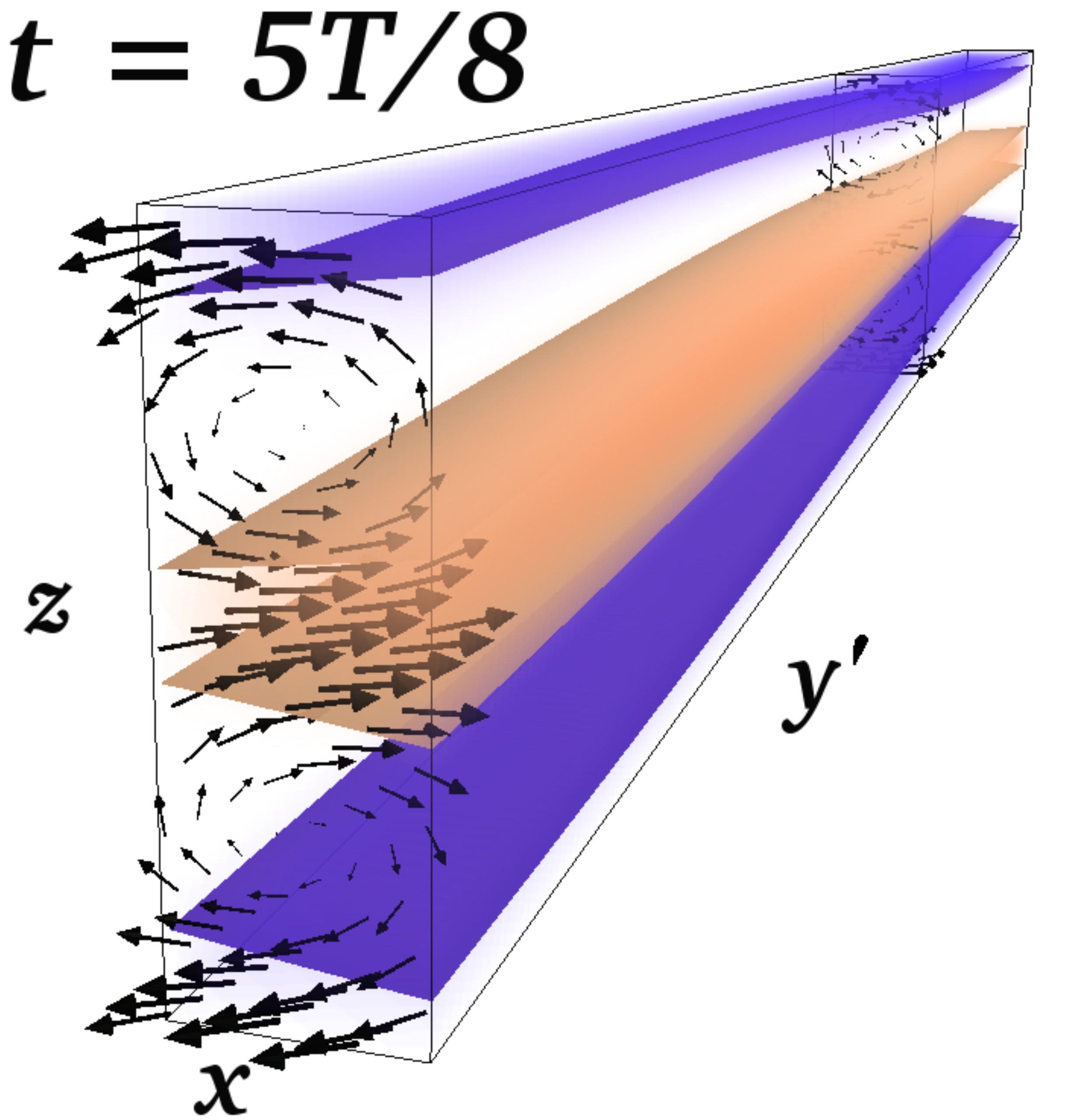}\includegraphics{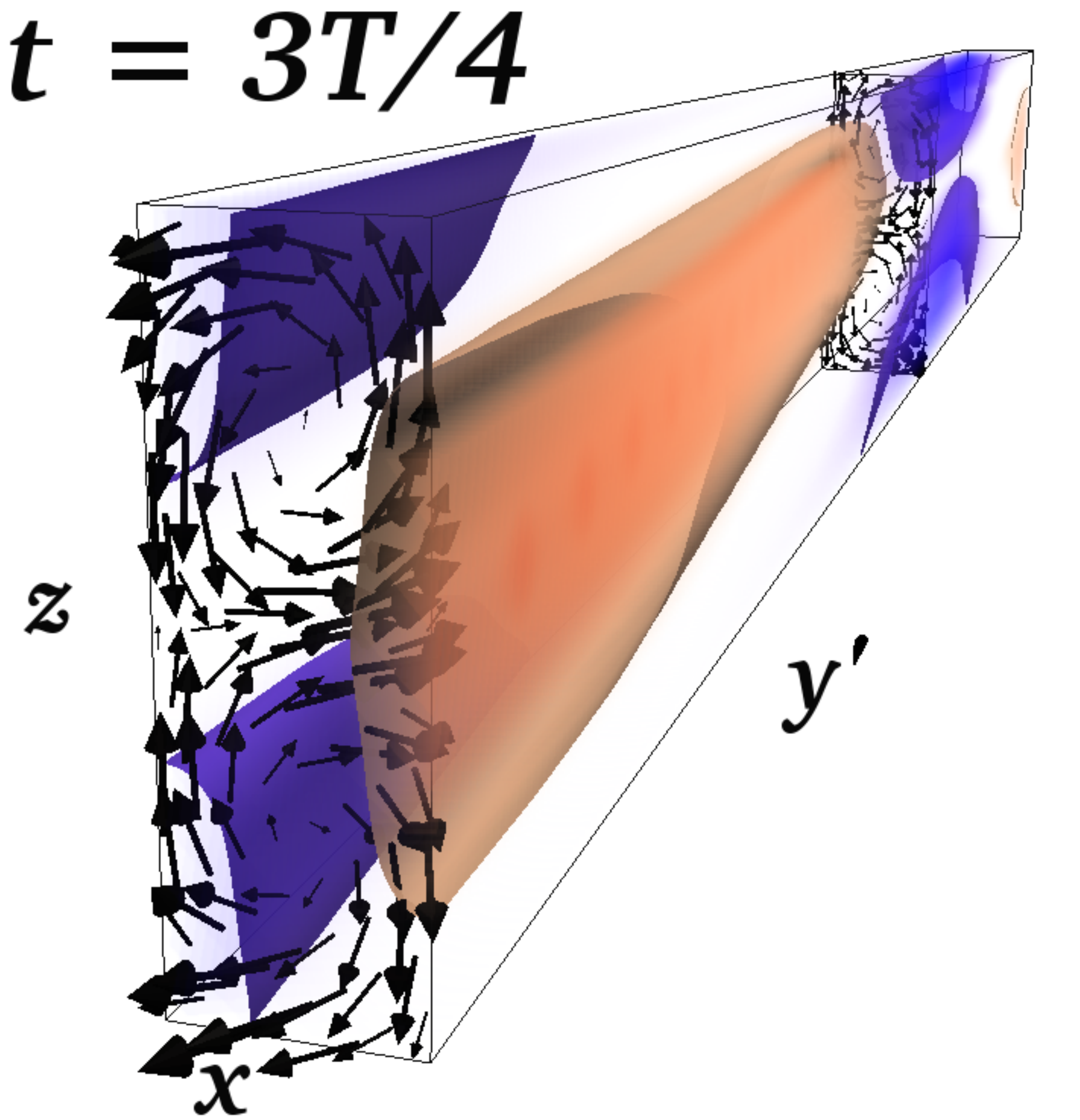}\includegraphics{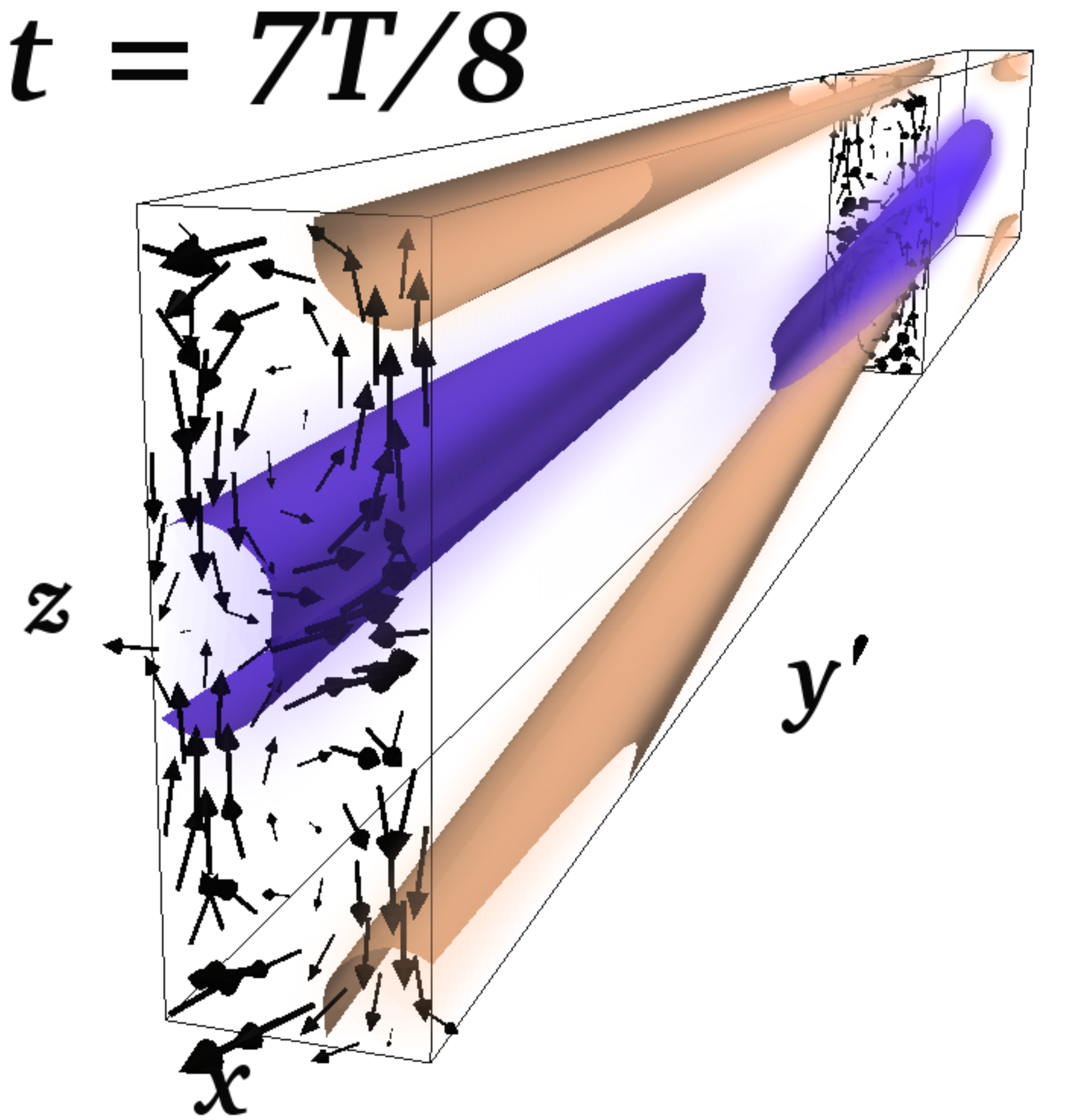}}
  \caption{(Color online) Volume renderings of $B_y$ and isosurfaces of $B=0.9$
    colored by $B_y$ every $T/8$ (positive $B_y$ in red/light gray, negative
    $B_y$ in blue-violet/dark gray). The coordinates of the bottom left corner of 
    each box are $x=0$, $y'=0$, $z=0$, where the $y'$ coordinate is 
    defined as $y'=y-(L_x/2) St$ (see Appendix).  The arrows field in 
    the $y'=0$ plane traces non-axisymmetric MRI velocity perturbations 
    (velocity perturbations in the $y'=L_y/2$ plane have opposite sign),
    whose effect is to distort and advect magnetic field lines of
    opposite polarities in opposite directions.}
\label{figure4}
\end{figure*}

\section{A nonlinear MRI dynamo cycle\label{cycle}}
Following the strategy defined in Sect.~\ref{frameworkstrategy}, we
then attempted to capture precisely the nonlinear cycle underlying
the recurrent dynamics spotted in Fig.~\ref{figure3} using the 
Newton-Krylov solver PEANUTS.

\subsection{Convergence with Newton's method}
Because of shearing-periodic boundary conditions, only cycles
whose period is a multiple of $L_y/(SL_x)$ are allowed in shearing boxes. 
The periodicity of the recurrent dynamics spotted in the DNS being very
close to $T=2\,L_y/(SL_x)\simeq 57.143\,S^{-1}$, this period was imposed
on the Newton solver. Initializing the solver with a snapshot of a DNS
displaying dynamical recurrences (the red full circle point of
Fig.~\ref{figure3}), we obtained convergence to a cycle after
$O(10)$ iterations with a relative error of $10^{-8}$. Each Newton
step requires $O(10)$ Krylov iterations to solve the Jacobian system
with a relative error smaller than $10^{-7}$. As each Krylov iteration
requires running a DNS over a cycle period, $O(100)$ numerical integrations 
are needed to ``capture'' this cycle with this iterative method.
The dynamics of this nonlinear MRI dynamo cycle is illustrated in
Fig.~\ref{figure4}. The results presented here are for a resolution
$(N_x,N_y,N_z)=(48,24,72)$ with 2/3 dealiasing (the same resolution
used to approach the cycle by DNS). Half that resolution already
ensures convergence to very good accuracy. Analyzing the energy
spectrum of the cycle, we only found differences of a few percent
between the full-resolution and half-resolution runs.

In respect of the observations made in Sect.~\ref{sensitive},
we emphasize that the cycle calculated by the means of Newton's method
is not a spurious numerical feature. The most important requirement
for convergence, as is standard with Newton's algorithm,
is that the DNS snapshot used as a starting guess for the algorithm be
in close-enough vicinity of the cycle.  Provided that this requirement
is satisfied, convergence to the cycle is robust and can be obtained by using (as a
starting guess for the Newton solver) various adequate DNS snapshots
resulting from integrations of various sets of initial conditions, for
various Courant-Friedrichs-Lewy conditions for the numerical
time-integrator, using either fixed or adaptative time-stepping.

Finally, it is important to stress that the cycle at hand is a genuine
solution to the full discretized nonlinear 3D MHD equations in the
shearing box, not just a solution to a low-dimensional, reduced
dynamical model of the problem. The imposition of symmetries in the
numerical resolution only helps to target nonlinear cycles with a
given natural symmetry (i.e. a symmetry allowed by the full MHD
equations in the geometry studied) more easily. In fact, once
convergence is achieved with the Newton solver, the cycle can be
integrated for several periods in a standard DNS without enforcing any
symmetry. 

\subsection{Description of the cycle}
The physics of the cycle is best understood by looking at
Fig.~\ref{figure4}. A large-scale axisymmetric toroidal field
$\overline{B}_y$ with $z$-dependent polarity dominates at $t=0$. The
MRI mediated by that field progressively amplifies weak, leading
non-axisymmetric perturbations. The effect of the instability
is to separate ``radially'' (in $x$) fluid particles initially attached to
individual, almost frozen-in field lines \citep{balbus98}, leading to their
non-axisymmetric distortion ($T/8$ to $T/4$). In addition, for the
cycle at hand, linear MRI velocity perturbations drag fields lines
with opposite polarities in opposite directions. As they get sheared,
MRI-amplified velocity perturbations take the form of a
non-axisymmetric pattern of counter-rotating poloidal flow cells. The
$z$ component of the flow advects distorted magnetic field lines with
opposite polarities in
opposite $z$ directions, effectively reversing regions of positive and
negative polarities ($T/4$). The $x$ component of the cellular flow 
eventually advects field lines back to their original $x$
location ($3T/8$), producing at $T/2$ an axisymmetric magnetic field
opposite to the original one. Hence, the reversal is the pure
outcome of the nonlinear evolution (self-interactions) of 
non-axisymmetric MRI perturbations. As a result of the confinement of 
MRI perturbations by the shearwise-modulated MRI-supporting toroidal
field, new leading perturbations seeds are generated during the first
half of the cycle. Their subsequent amplification and nonlinear
evolution  during the second half of the cycle result in a new field
reversal after another $T/2$, back to the initial state.  

\begin{figure}[t]
\resizebox{\hsize}{!}{\includegraphics{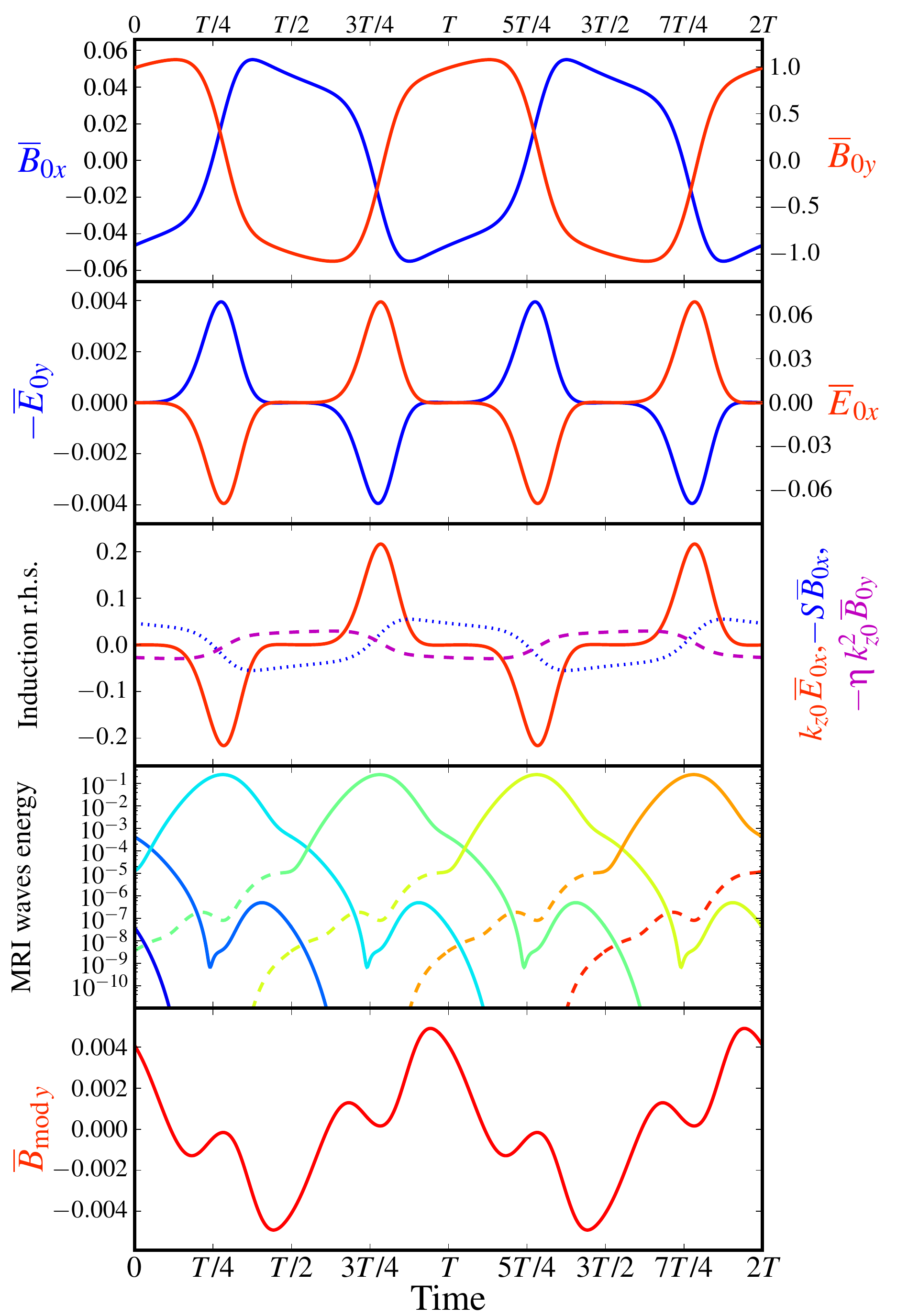}}
  \caption{(Color online) From top to bottom: evolution over two cycles of a)
    $\overline{B}_{0x}$ (blue/dark gray) and $\overline{B}_{0y}$  (red/light gray) ;
    b) $-\overline{E}_{0y}$ (blue/dark gray, source term for $\overline{B}_{0x}$) and
    $\overline{E}_{0x}$ (red / light gray, source term for $\overline{B}_{0y}$) ; c)
    amplitudes of the three terms on the r.h.s. of the projection of
    the axisymmetric induction Eq.~(\ref{eq:inducB0}) on $\vec{e}_y$
    (dotted blue: $\Omega$ effect, full line red: nonlinear EMF,
    dashed magenta: magnetic diffusion) ; d) total energy of
    successive MRI shearing waves with $k_y=\pm
    2\pi/L_y$ (rainbow colors, dashed/full line: leading/trailing
    phase. Each color represents a different shearing wave) ; e),
    amplitude of the axisymmetric magnetic field modulation
    $\overline{B}_{\mathrm{mod}\,y}$.}
\label{figure5}
\end{figure}

This qualitative physical scenario  is fully supported by a
quantitative analysis of the time-evolution
(Fig.~\ref{figure5}). The li\-near MRI of the
$z$-dependent axi\-symmetric toroi\-dal field $\overline{B}_{0y}$
(Fig.~\ref{figure5}a in red) amplifies successive non-axisymmetric
shearing MRI waves (Fig.~\ref{figure5}d, rainbow colors) with $k_y=\pm
2\pi/L_y$ and a $k_z$ spectrum. The nonlinear self-interaction of a
single wave packet translates into an axisymmetric EMF
$\overline{\vec{E}}_0$ whose $x$ and $y$ components are clearly
responsible for the reversals of $\overline{\vec{B}}_0$
(Fig.~\ref{figure5}b). Interestingly, a calculation of the various
terms on the r.h.s. of the projection of Eq.~(\ref{eq:inducB0}) on
$\vec{e}_y$ throughout the cycle shows that the $\Omega$ effect is not
the dominant  term in this equation for this particular cycle and
geometry (Fig.~\ref{figure5}c). Hence, $\overline{B}_{0x}$ and
$\overline{B}_{0y}$ are almost in antiphase, whereas they are almost
in quadrature for the pseudo-cycle described in
Ref.~\cite{lesur08}. Fig.~\ref{figure5}d clearly demonstrates the
periodic seeding of new leading perturbations (yellow, green, light
blue etc. represent successive shearing waves, as in
Fig.~\ref{figure2}). The evolution of the amplitude of
$\overline{B}_{\mathrm{mod}\,y}$, defined in Eq.~(\ref{eq:bmed}),
is depicted in Fig.~\ref{figure5}e. $\overline{B}_{\mathrm{mod}\,y}$
is very small compared to $\overline{B}_{0y}$, which demonstrates 
that a very weak confinement of the MRI is actually sufficient to
generate new leading waves.

As seen in Fig.~\ref{figure4}, the reason why such a cycle can be
computed at fairly low resolutions (in particular in the $y$ direction)
is that it is a large-scale, time-dependent coherent structure whose
non-axisymmetric dynamics (and the corresponding axisymmetric dynamo
feedback $\overline{\vec{E}}_0$ that it induces) is largely dominated
by MRI shearing waves supported by the fundamental Fourier mode
$k_y=2\pi/L_y$ of the shearing box in the $y$ direction. We also
note that all the physical processes involved are not actually
specific to the shearing box, and may therefore also be present in
cylindrical geometry, with non-axisymmetric perturbations taking on 
the form of sheared spiral waves. 

Finally, remark that the energy required to sustain
the three-dimensional cyclic dynamics against resistive and viscous
dissipation is extracted from the shear (which is the only available
energy source of the system) thanks to the toroidal MRI of
non-axisymmetric shearing waves. 

\subsection{Stability}
Using the Floquet eigenvalue solver, we found that the cycle has a
single unstable eigenmode, with a Floquet eigenvalue $\Lambda\simeq
32.33$ corresponding to a positive growth rate
$\lambda=\ln\Lambda/T\simeq 0.061\, S$ (at half resolution,
$\Lambda=22.63$, $\lambda\simeq 0.055\, S$), and the same
$\mathcal{A}_1$ symmetry. Hence, any small
perturbation to the cycle, independently of its initial amplitude,
ultimately kicks the system out of its initially periodic trajectory
in phase-space. This exponential instability of the cycle was
observed in the simulations and leads to complete escape from the
neighbourhood of the cyclic solution after a few hundred shearing
times. Unstable periodic orbits such as this one are a typical feature
of chaotic dynamical systems and play an important role in their
dynamics (see discussion in Sect.~\ref{discussion}).

\subsection{Investigating the nature of dynamo action}
\begin{figure}[t]
\resizebox{\hsize}{!}{\includegraphics{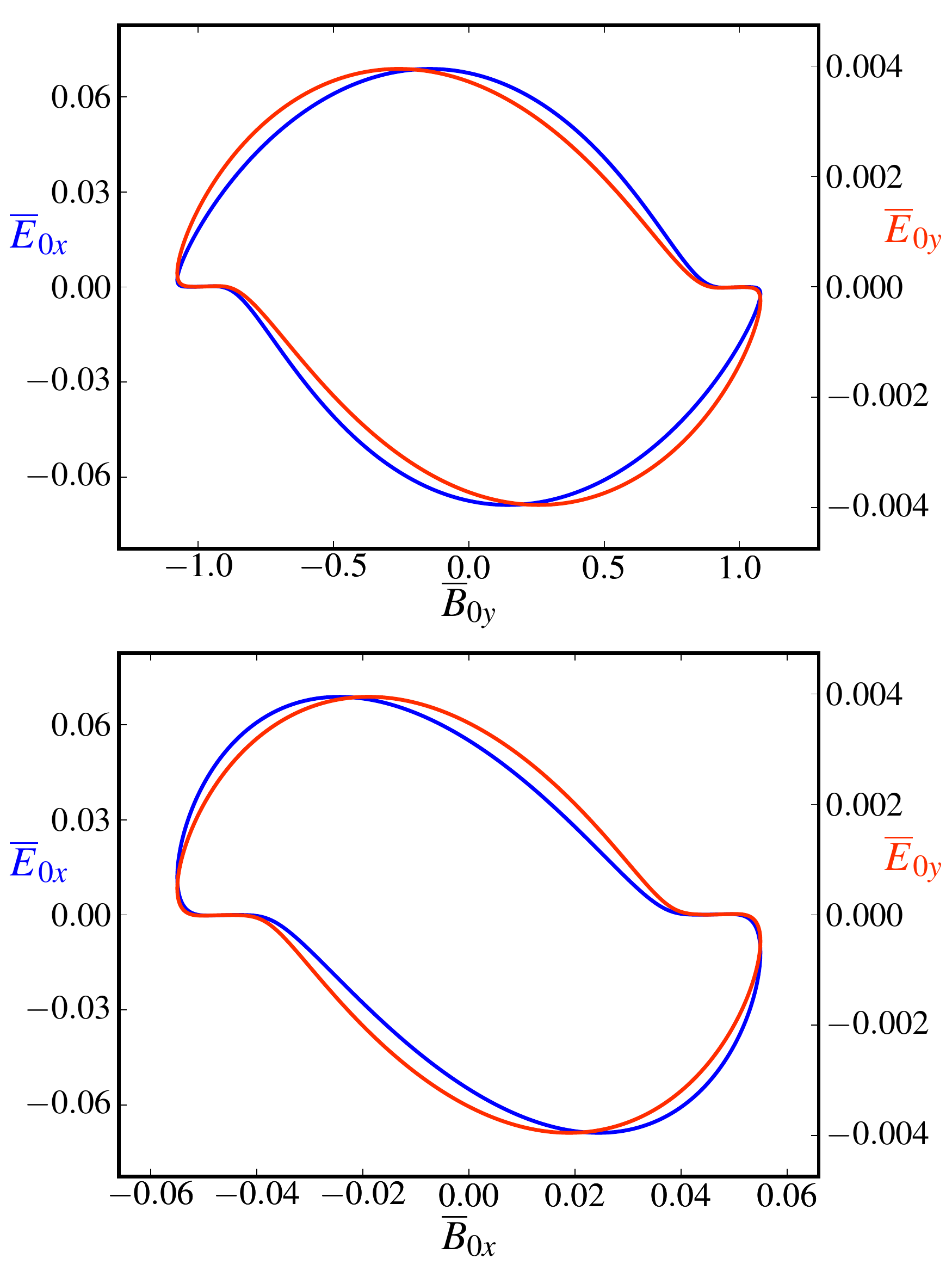}}
  \caption{(Color online) Projection of the periodic MRI dynamo orbit
in the $\overline{E}_x$ (blue/dark gray) | $\overline{E}_y$ (red/light 
gray) vs.  $\overline{B}_x$ | $\overline{B}_y$ planes.}
\label{figure6}
\end{figure}

One may finally wonder if the nonlinear couplings leading to this
dynamo can be described by standard mean-field theory. $\alpha^2$
or $\alpha\Omega$ dynamos are ruled out: both net kinetic and current
helicities are negligible in the DNS. For our cycle, the axisymmetric
field and EMF are largely dominated by their projection
$\overline{\vec{B}}_0$ and $\overline{\vec{E}}_0$ on $\cos (2\pi
z/L_z)$ and $\sin (2\pi z/L_z)$ planforms (see
Eqs.~(\ref{eq:axifieldfundamental})-(\ref{eq:EMFfundamental})), 
so one may be tempted to interpret the dynamo feedback in terms
of a  non-diagonal ``turbulent resistivity'' $\Bar{\Bar{\eta}}$ tensor
\citep{gressel10} that would couple the amplitudes of the components
of $\overline{\vec{B}}_0$ and $\overline{\vec{E}}_0$ linearly. However,
not only is the instantaneous relationship between these two quantities 
unambiguously nonlinear (Fig.~\ref{figure6}), it also looks rather
difficult to model analytically based on simple physical arguments.
Hence, this periodic MRI dynamo does not reduce to a simple standard
mean-field dynamo, a conclusion that also seems to apply to the
magnetic buoyancy-driven dynamo \citep{cline03,davies10}.

\section{Discussion and conclusions\label{discussion}}
We have presented the first accurate numerical determination of a
cyclic nonlinear MRI dynamo solution to the original 3D dissipative
incompressible MHD equations in the regime of Keplerian differential
rotation. Preliminary investigations indicate that its existence 
is not limited to a narrow range of $\rey$ and $\reym$, unlike the
plane Couette flow stationary solution reported in
Ref.~\cite{rincon07b}, and may notably extend down to low magnetic
Prandtl number regimes in which recent studies have found it difficult
to obtain sustained MRI dynamo turbulence \citep{fromang07b}. A
detailed parametric study of the cycle and an assessment of its
importance for the MRI dynamo transition problem is currently underway
and will be presented in a separate paper.

Overall, the discovery of this MRI dynamo limit cycle significantly
extends and consolidates earlier findings \citep{rincon07b}  and claims
\citep{rincon08,rempel10} that dynamo action and MHD turbulence in
shear flows prone to local MHD instabilities has a similar nature to
hydrodynamic turbulence in shear flows. The results notably provide
clear evidence that magnetic field generation
and sustenance at moderate $\rey$, $\reym$ in a numerical set-up
comparable to that of most MRI dynamo simulations so far results from
a nonlinear mechanism of interaction between axisymmetric
fields and non-axisymmetric instability modes. The detailed
description of the process may guide future studies of dynamos
mediated by MHD instabilities other than the
MRI \citep{spruit02,cline03,braithwaite06,miesch07,tobias11}.

Our findings also demonstrate that new theoretical models
of large-scale dynamo action are required. Progress on this problem
may be possible thanks to periodic orbit theory \citep{cvit92}, which
offers formal mathematical connections between  the statistical
properties of dynamical systems with a small number of active degrees
of freedom and their unstable cycles. One should therefore
attempt to identify new MRI dynamo cycles, analyze their dependence on
various parameters and assess their relative dynamical importance
using stability analysis. It is almost certain that,
similarly to hydrodynamic shear flows \citep{viswanath07},
many such structures exist in the phase space of MHD-unstable shear
flows at moderate and large $\rey$ and $\reym$.  

This approach may also prove useful for the astrophysics problem of
turbulent transport of angular momentum in accretion disks
\citep{balbus98}. The fact that the MRI dynamo cycle described above
generates an average turbulent MHD stress $\alpha\sim 0.025\, (SL)^2$
directly comparable to that obtained in direct numerical simulations
of similar configurations \citep{fromang07b} constitutes a notable
preliminary finding in this respect.

We finally point out that the MHD findings reported in this study,
which have partly been inspired by recent advances in the field of
shear flow hydrodynamics, may in return be helpful to make progress on 
understanding hydrodynamic shear flow turbulence. The leading wave
regeneration mechanism that was clearly identified in the shearing box
in this work offers an interesting connexion between three-dimensional
turbulence in wall-bounded and homogeneous shear flows, which may
notably help to understand if regeneration pseudo-cycles in the latter
(see e.g. Ref.~\citep{gualtieri02}) share a common physical origin
with nonlinear cycles in wall-bounded shear flows \citep{viswanath07}.

\acknowledgments
This research was supported in part by the National Science Foundation
under Grant No. PHY05-51164, by the Leverhulme Trust Network for
Magnetized Plasma Turbulence and by the French National Program
for Stellar Physics (PNPS). Calculations were carried out on the
CALMIP supercomputer (CICT, University of Toulouse). We would like to
thank Michael Proctor, Erico Rempel, S\'ebastien Fromang, Alexander
Schekochihin and Tobias Heinemann for several useful discussions on
the problem.

\bibliographystyle{apsrev}
\bibliography{pre_herault}

\begin{thebibliography}{58}
\expandafter\ifx\csname natexlab\endcsname\relax\def\natexlab#1{#1}\fi
\expandafter\ifx\csname bibnamefont\endcsname\relax
  \def\bibnamefont#1{#1}\fi
\expandafter\ifx\csname bibfnamefont\endcsname\relax
  \def\bibfnamefont#1{#1}\fi
\expandafter\ifx\csname citenamefont\endcsname\relax
  \def\citenamefont#1{#1}\fi
\expandafter\ifx\csname url\endcsname\relax
  \def\url#1{\texttt{#1}}\fi
\expandafter\ifx\csname urlprefix\endcsname\relax\def\urlprefix{URL }\fi
\providecommand{\bibinfo}[2]{#2}
\providecommand{\eprint}[2][]{\url{#2}}

\bibitem[{\citenamefont{{Moffatt}}(1977)}]{moffatt77}
\bibinfo{author}{\bibfnamefont{H.~K.} \bibnamefont{{Moffatt}}},
  \emph{\bibinfo{title}{{Magnetic field generation in electrically conducting
  fluids.}}} (\bibinfo{publisher}{Cambridge University Press},
  \bibinfo{year}{1977}).

\bibitem[{\citenamefont{{Parker}}(1955)}]{parker55}
\bibinfo{author}{\bibfnamefont{E.~N.} \bibnamefont{{Parker}}},
  \bibinfo{journal}{ApJ} \textbf{\bibinfo{volume}{122}}, \bibinfo{pages}{293}
  (\bibinfo{year}{1955}).

\bibitem[{\citenamefont{{Brandenburg} and {Subramanian}}(2005)}]{branden05}
\bibinfo{author}{\bibfnamefont{A.}~\bibnamefont{{Brandenburg}}}
  \bibnamefont{and}
  \bibinfo{author}{\bibfnamefont{K.}~\bibnamefont{{Subramanian}}},
  \bibinfo{journal}{Phys. Rep.} \textbf{\bibinfo{volume}{417}},
  \bibinfo{pages}{1} (\bibinfo{year}{2005}).

\bibitem[{\citenamefont{{Brandenburg} et~al.}(1995)\citenamefont{{Brandenburg},
  {Nordlund}, {Stein}, and {Torkelsson}}}]{branden95}
\bibinfo{author}{\bibfnamefont{A.}~\bibnamefont{{Brandenburg}}},
  \bibinfo{author}{\bibfnamefont{A.}~\bibnamefont{{Nordlund}}},
  \bibinfo{author}{\bibfnamefont{R.~F.} \bibnamefont{{Stein}}},
  \bibnamefont{and}
  \bibinfo{author}{\bibfnamefont{U.}~\bibnamefont{{Torkelsson}}},
  \bibinfo{journal}{ApJ} \textbf{\bibinfo{volume}{446}}, \bibinfo{pages}{741}
  (\bibinfo{year}{1995}).

\bibitem[{\citenamefont{{Hawley} et~al.}(1996)\citenamefont{{Hawley}, {Gammie},
  and {Balbus}}}]{hawley96}
\bibinfo{author}{\bibfnamefont{J.~F.} \bibnamefont{{Hawley}}},
  \bibinfo{author}{\bibfnamefont{C.~F.} \bibnamefont{{Gammie}}},
  \bibnamefont{and} \bibinfo{author}{\bibfnamefont{S.~A.}
  \bibnamefont{{Balbus}}}, \bibinfo{journal}{ApJ}
  \textbf{\bibinfo{volume}{464}}, \bibinfo{pages}{690} (\bibinfo{year}{1996}).

\bibitem[{\citenamefont{{Drecker} et~al.}(2000)\citenamefont{{Drecker},
  {R{\"u}diger}, and {Hollerbach}}}]{drecker2000}
\bibinfo{author}{\bibfnamefont{A.}~\bibnamefont{{Drecker}}},
  \bibinfo{author}{\bibfnamefont{G.}~\bibnamefont{{R{\"u}diger}}},
  \bibnamefont{and}
  \bibinfo{author}{\bibfnamefont{R.}~\bibnamefont{{Hollerbach}}},
  \bibinfo{journal}{MNRAS} \textbf{\bibinfo{volume}{317}}, \bibinfo{pages}{45}
  (\bibinfo{year}{2000}).

\bibitem[{\citenamefont{{Spruit}}(2002)}]{spruit02}
\bibinfo{author}{\bibfnamefont{H.~C.} \bibnamefont{{Spruit}}},
  \bibinfo{journal}{A{\&}A} \textbf{\bibinfo{volume}{381}},
  \bibinfo{pages}{923} (\bibinfo{year}{2002}).

\bibitem[{\citenamefont{{Cline} et~al.}(2003)\citenamefont{{Cline}, {Brummell},
  and {Cattaneo}}}]{cline03}
\bibinfo{author}{\bibfnamefont{K.~S.} \bibnamefont{{Cline}}},
  \bibinfo{author}{\bibfnamefont{N.~H.} \bibnamefont{{Brummell}}},
  \bibnamefont{and}
  \bibinfo{author}{\bibfnamefont{F.}~\bibnamefont{{Cattaneo}}},
  \bibinfo{journal}{ApJ} \textbf{\bibinfo{volume}{599}}, \bibinfo{pages}{1449}
  (\bibinfo{year}{2003}).

\bibitem[{\citenamefont{{Braithwaite}}(2006)}]{braithwaite06}
\bibinfo{author}{\bibfnamefont{J.}~\bibnamefont{{Braithwaite}}},
  \bibinfo{journal}{A{\&}A} \textbf{\bibinfo{volume}{449}},
  \bibinfo{pages}{451} (\bibinfo{year}{2006}).

\bibitem[{\citenamefont{{Rincon} et~al.}(2007)\citenamefont{{Rincon},
  {Ogilvie}, and {Proctor}}}]{rincon07b}
\bibinfo{author}{\bibfnamefont{F.}~\bibnamefont{{Rincon}}},
  \bibinfo{author}{\bibfnamefont{G.~I.} \bibnamefont{{Ogilvie}}},
  \bibnamefont{and} \bibinfo{author}{\bibfnamefont{M.~R.~E.}
  \bibnamefont{{Proctor}}}, \bibinfo{journal}{Phys. Rev. Lett.}
  \textbf{\bibinfo{volume}{98}}, \bibinfo{pages}{254502}
  (\bibinfo{year}{2007}).

\bibitem[{\citenamefont{{Lesur} and {Ogilvie}}(2008{\natexlab{a}})}]{lesur08}
\bibinfo{author}{\bibfnamefont{G.}~\bibnamefont{{Lesur}}} \bibnamefont{and}
  \bibinfo{author}{\bibfnamefont{G.~I.} \bibnamefont{{Ogilvie}}},
  \bibinfo{journal}{A{\&}A} \textbf{\bibinfo{volume}{488}},
  \bibinfo{pages}{451} (\bibinfo{year}{2008}{\natexlab{a}}).

\bibitem[{\citenamefont{{Davis} et~al.}(2010)\citenamefont{{Davis}, {Stone},
  and {Pessah}}}]{davis10}
\bibinfo{author}{\bibfnamefont{S.~W.} \bibnamefont{{Davis}}},
  \bibinfo{author}{\bibfnamefont{J.~M.} \bibnamefont{{Stone}}},
  \bibnamefont{and} \bibinfo{author}{\bibfnamefont{M.~E.}
  \bibnamefont{{Pessah}}}, \bibinfo{journal}{ApJ}
  \textbf{\bibinfo{volume}{713}}, \bibinfo{pages}{52} (\bibinfo{year}{2010}).

\bibitem[{\citenamefont{{Tobias} et~al.}(2011)\citenamefont{{Tobias},
  {Cattaneo}, and {Brummell}}}]{tobias11}
\bibinfo{author}{\bibfnamefont{S.~M.} \bibnamefont{{Tobias}}},
  \bibinfo{author}{\bibfnamefont{F.}~\bibnamefont{{Cattaneo}}},
  \bibnamefont{and} \bibinfo{author}{\bibfnamefont{N.~H.}
  \bibnamefont{{Brummell}}}, \bibinfo{journal}{ApJ}
  \textbf{\bibinfo{volume}{728}}, \bibinfo{pages}{153} (\bibinfo{year}{2011}).

\bibitem[{\citenamefont{{Rincon} et~al.}(2008)\citenamefont{{Rincon},
  {Ogilvie}, {Proctor}, and {Cossu}}}]{rincon08}
\bibinfo{author}{\bibfnamefont{F.}~\bibnamefont{{Rincon}}},
  \bibinfo{author}{\bibfnamefont{G.~I.} \bibnamefont{{Ogilvie}}},
  \bibinfo{author}{\bibfnamefont{M.~R.~E.} \bibnamefont{{Proctor}}},
  \bibnamefont{and} \bibinfo{author}{\bibfnamefont{C.}~\bibnamefont{{Cossu}}},
  \bibinfo{journal}{Astron. Nachr.} \textbf{\bibinfo{volume}{329}},
  \bibinfo{pages}{750} (\bibinfo{year}{2008}).

\bibitem[{\citenamefont{{Rempel} et~al.}(2010)\citenamefont{{Rempel}, {Lesur},
  and {Proctor}}}]{rempel10}
\bibinfo{author}{\bibfnamefont{E.~L.} \bibnamefont{{Rempel}}},
  \bibinfo{author}{\bibfnamefont{G.}~\bibnamefont{{Lesur}}}, \bibnamefont{and}
  \bibinfo{author}{\bibfnamefont{M.~R.~E.} \bibnamefont{{Proctor}}},
  \bibinfo{journal}{Phys. Rev. Lett.} \textbf{\bibinfo{volume}{105}},
  \bibinfo{pages}{044501} (\bibinfo{year}{2010}).

\bibitem[{\citenamefont{{Hamilton} et~al.}(1995)\citenamefont{{Hamilton},
  {Kim}, and {Waleffe}}}]{hamilton95}
\bibinfo{author}{\bibfnamefont{J.~M.} \bibnamefont{{Hamilton}}},
  \bibinfo{author}{\bibfnamefont{J.}~\bibnamefont{{Kim}}}, \bibnamefont{and}
  \bibinfo{author}{\bibfnamefont{F.}~\bibnamefont{{Waleffe}}},
  \bibinfo{journal}{J. Fluid Mech.} \textbf{\bibinfo{volume}{287}},
  \bibinfo{pages}{317} (\bibinfo{year}{1995}).

\bibitem[{\citenamefont{{Waleffe}}(1997)}]{waleffe97}
\bibinfo{author}{\bibfnamefont{F.}~\bibnamefont{{Waleffe}}},
  \bibinfo{journal}{Phys. Fluids} \textbf{\bibinfo{volume}{9}},
  \bibinfo{pages}{883} (\bibinfo{year}{1997}).

\bibitem[{\citenamefont{{Faisst} and {Eckhardt}}(2004)}]{faisst04}
\bibinfo{author}{\bibfnamefont{H.}~\bibnamefont{{Faisst}}} \bibnamefont{and}
  \bibinfo{author}{\bibfnamefont{B.}~\bibnamefont{{Eckhardt}}},
  \bibinfo{journal}{J. Fluid Mech.} \textbf{\bibinfo{volume}{504}},
  \bibinfo{pages}{343} (\bibinfo{year}{2004}).

\bibitem[{\citenamefont{{Hof} et~al.}(2006)\citenamefont{{Hof}, {Westerweel},
  {Schneider}, and {Eckhardt}}}]{hof06}
\bibinfo{author}{\bibfnamefont{B.}~\bibnamefont{{Hof}}},
  \bibinfo{author}{\bibfnamefont{J.}~\bibnamefont{{Westerweel}}},
  \bibinfo{author}{\bibfnamefont{T.~M.} \bibnamefont{{Schneider}}},
  \bibnamefont{and}
  \bibinfo{author}{\bibfnamefont{B.}~\bibnamefont{{Eckhardt}}},
  \bibinfo{journal}{Nature} \textbf{\bibinfo{volume}{443}}, \bibinfo{pages}{59}
  (\bibinfo{year}{2006}).

\bibitem[{\citenamefont{{Eckhardt} et~al.}(2007)\citenamefont{{Eckhardt},
  {Schneider}, {Hof}, and {Westerweel}}}]{eckhardt07}
\bibinfo{author}{\bibfnamefont{B.}~\bibnamefont{{Eckhardt}}},
  \bibinfo{author}{\bibfnamefont{T.~M.} \bibnamefont{{Schneider}}},
  \bibinfo{author}{\bibfnamefont{B.}~\bibnamefont{{Hof}}}, \bibnamefont{and}
  \bibinfo{author}{\bibfnamefont{J.}~\bibnamefont{{Westerweel}}},
  \bibinfo{journal}{Ann. Rev. Fluid Mech.} \textbf{\bibinfo{volume}{39}},
  \bibinfo{pages}{447} (\bibinfo{year}{2007}).

\bibitem[{\citenamefont{{Waleffe}}(1998)}]{waleffe98}
\bibinfo{author}{\bibfnamefont{F.}~\bibnamefont{{Waleffe}}},
  \bibinfo{journal}{Phys. Rev. Lett.} \textbf{\bibinfo{volume}{81}},
  \bibinfo{pages}{4140} (\bibinfo{year}{1998}).

\bibitem[{\citenamefont{{Faisst} and {Eckhardt}}(2003)}]{faisst03}
\bibinfo{author}{\bibfnamefont{H.}~\bibnamefont{{Faisst}}} \bibnamefont{and}
  \bibinfo{author}{\bibfnamefont{B.}~\bibnamefont{{Eckhardt}}},
  \bibinfo{journal}{Phys. Rev. Lett.} \textbf{\bibinfo{volume}{91}},
  \bibinfo{pages}{224502} (\bibinfo{year}{2003}).

\bibitem[{\citenamefont{{Wedin} and {Kerswell}}(2004)}]{wedin04}
\bibinfo{author}{\bibfnamefont{H.}~\bibnamefont{{Wedin}}} \bibnamefont{and}
  \bibinfo{author}{\bibfnamefont{R.~R.} \bibnamefont{{Kerswell}}},
  \bibinfo{journal}{J. Fluid Mech.} \textbf{\bibinfo{volume}{508}},
  \bibinfo{pages}{333} (\bibinfo{year}{2004}).

\bibitem[{\citenamefont{{Gibson} et~al.}(2009)\citenamefont{{Gibson},
  {Halcrow}, and {Cvitanovi{\'c}}}}]{gibson09}
\bibinfo{author}{\bibfnamefont{J.~F.} \bibnamefont{{Gibson}}},
  \bibinfo{author}{\bibfnamefont{J.}~\bibnamefont{{Halcrow}}},
  \bibnamefont{and}
  \bibinfo{author}{\bibfnamefont{P.}~\bibnamefont{{Cvitanovi{\'c}}}},
  \bibinfo{journal}{J. Fluid Mech.} \textbf{\bibinfo{volume}{638}},
  \bibinfo{pages}{243} (\bibinfo{year}{2009}).

\bibitem[{\citenamefont{{Kawahara} and {Kida}}(2001)}]{kawahara01}
\bibinfo{author}{\bibfnamefont{G.}~\bibnamefont{{Kawahara}}} \bibnamefont{and}
  \bibinfo{author}{\bibfnamefont{S.}~\bibnamefont{{Kida}}},
  \bibinfo{journal}{J. Fluid Mech.} \textbf{\bibinfo{volume}{449}},
  \bibinfo{pages}{291} (\bibinfo{year}{2001}).

\bibitem[{\citenamefont{{Toh} and {Itano}}(2003)}]{toh03}
\bibinfo{author}{\bibfnamefont{S.}~\bibnamefont{{Toh}}} \bibnamefont{and}
  \bibinfo{author}{\bibfnamefont{T.}~\bibnamefont{{Itano}}},
  \bibinfo{journal}{J. Fluid Mech.} \textbf{\bibinfo{volume}{481}},
  \bibinfo{pages}{67} (\bibinfo{year}{2003}).

\bibitem[{\citenamefont{{Viswanath}}(2007)}]{viswanath07}
\bibinfo{author}{\bibfnamefont{D.}~\bibnamefont{{Viswanath}}},
  \bibinfo{journal}{J. Fluid Mech.} \textbf{\bibinfo{volume}{580}},
  \bibinfo{pages}{339} (\bibinfo{year}{2007}).

\bibitem[{\citenamefont{{Halcrow} et~al.}(2009)\citenamefont{{Halcrow},
  {Gibson}, {Cvitanovi{\'c}}, and {Viswanath}}}]{halcrow09}
\bibinfo{author}{\bibfnamefont{J.}~\bibnamefont{{Halcrow}}},
  \bibinfo{author}{\bibfnamefont{J.~F.} \bibnamefont{{Gibson}}},
  \bibinfo{author}{\bibfnamefont{P.}~\bibnamefont{{Cvitanovi{\'c}}}},
  \bibnamefont{and}
  \bibinfo{author}{\bibfnamefont{D.}~\bibnamefont{{Viswanath}}},
  \bibinfo{journal}{J. Fluid Mech.} p. \bibinfo{pages}{365}
  (\bibinfo{year}{2009}).

\bibitem[{\citenamefont{{Balbus} and {Hawley}}(1998)}]{balbus98}
\bibinfo{author}{\bibfnamefont{S.~A.} \bibnamefont{{Balbus}}} \bibnamefont{and}
  \bibinfo{author}{\bibfnamefont{J.~F.} \bibnamefont{{Hawley}}},
  \bibinfo{journal}{Rev. Mod. Phys.} \textbf{\bibinfo{volume}{70}},
  \bibinfo{pages}{1} (\bibinfo{year}{1998}).

\bibitem[{\citenamefont{{Fromang} et~al.}(2007)\citenamefont{{Fromang},
  {Papaloizou}, {Lesur}, and {Heinemann}}}]{fromang07b}
\bibinfo{author}{\bibfnamefont{S.}~\bibnamefont{{Fromang}}},
  \bibinfo{author}{\bibfnamefont{J.}~\bibnamefont{{Papaloizou}}},
  \bibinfo{author}{\bibfnamefont{G.}~\bibnamefont{{Lesur}}}, \bibnamefont{and}
  \bibinfo{author}{\bibfnamefont{T.}~\bibnamefont{{Heinemann}}},
  \bibinfo{journal}{A{\&}A} \textbf{\bibinfo{volume}{476}},
  \bibinfo{pages}{1123} (\bibinfo{year}{2007}).

\bibitem[{\citenamefont{{Goldreich} and {Lynden-Bell}}(1965)}]{goldreich65}
\bibinfo{author}{\bibfnamefont{P.}~\bibnamefont{{Goldreich}}} \bibnamefont{and}
  \bibinfo{author}{\bibfnamefont{D.}~\bibnamefont{{Lynden-Bell}}},
  \bibinfo{journal}{MNRAS} \textbf{\bibinfo{volume}{130}}, \bibinfo{pages}{125}
  (\bibinfo{year}{1965}).

\bibitem[{\citenamefont{{Fleming} et~al.}(2000)\citenamefont{{Fleming},
  {Stone}, and {Hawley}}}]{fleming00}
\bibinfo{author}{\bibfnamefont{T.~P.} \bibnamefont{{Fleming}}},
  \bibinfo{author}{\bibfnamefont{J.~M.} \bibnamefont{{Stone}}},
  \bibnamefont{and} \bibinfo{author}{\bibfnamefont{J.~F.}
  \bibnamefont{{Hawley}}}, \bibinfo{journal}{ApJ}
  \textbf{\bibinfo{volume}{530}}, \bibinfo{pages}{464} (\bibinfo{year}{2000}).

\bibitem[{\citenamefont{{Pumir}}(1996)}]{pumir96}
\bibinfo{author}{\bibfnamefont{A.}~\bibnamefont{{Pumir}}},
  \bibinfo{journal}{Phys. Fluids} \textbf{\bibinfo{volume}{8}},
  \bibinfo{pages}{3112} (\bibinfo{year}{1996}).

\bibitem[{\citenamefont{{Gualtieri} et~al.}(2002)\citenamefont{{Gualtieri},
  {Casciola}, {Benzi}, {Amati}, and {Piva}}}]{gualtieri02}
\bibinfo{author}{\bibfnamefont{P.}~\bibnamefont{{Gualtieri}}},
  \bibinfo{author}{\bibfnamefont{C.~M.} \bibnamefont{{Casciola}}},
  \bibinfo{author}{\bibfnamefont{R.}~\bibnamefont{{Benzi}}},
  \bibinfo{author}{\bibfnamefont{G.}~\bibnamefont{{Amati}}}, \bibnamefont{and}
  \bibinfo{author}{\bibfnamefont{R.}~\bibnamefont{{Piva}}},
  \bibinfo{journal}{Phys. Fluids} \textbf{\bibinfo{volume}{14}},
  \bibinfo{pages}{583} (\bibinfo{year}{2002}).

\bibitem[{\citenamefont{{Lesur} and {Longaretti}}(2007)}]{lesur07}
\bibinfo{author}{\bibfnamefont{G.}~\bibnamefont{{Lesur}}} \bibnamefont{and}
  \bibinfo{author}{\bibfnamefont{P.-Y.} \bibnamefont{{Longaretti}}},
  \bibinfo{journal}{MNRAS} \textbf{\bibinfo{volume}{378}},
  \bibinfo{pages}{1471} (\bibinfo{year}{2007}).

\bibitem[{\citenamefont{{Umurhan} and {Regev}}(2004)}]{umurhan04}
\bibinfo{author}{\bibfnamefont{O.~M.} \bibnamefont{{Umurhan}}}
  \bibnamefont{and} \bibinfo{author}{\bibfnamefont{O.}~\bibnamefont{{Regev}}},
  \bibinfo{journal}{A{\&}A} \textbf{\bibinfo{volume}{427}},
  \bibinfo{pages}{855} (\bibinfo{year}{2004}).

\bibitem[{\citenamefont{Thomson (Lord~Kelvin)}(1887)}]{kelvin1887}
\bibinfo{author}{\bibfnamefont{W.}~\bibnamefont{Thomson (Lord~Kelvin)}},
  \bibinfo{journal}{{Phil. Mag.}} \textbf{\bibinfo{volume}{24}},
  \bibinfo{pages}{188} (\bibinfo{year}{1887}).

\bibitem[{\citenamefont{{Orr}}(1907)}]{orr1907}
\bibinfo{author}{\bibfnamefont{W.~M.} \bibnamefont{{Orr}}},
  \bibinfo{journal}{{Proc. R. Irish Acad. A.}} \textbf{\bibinfo{volume}{27}},
  \bibinfo{pages}{9} (\bibinfo{year}{1907}).

\bibitem[{\citenamefont{{Knobloch}}(1985)}]{knobloch85}
\bibinfo{author}{\bibfnamefont{E.}~\bibnamefont{{Knobloch}}},
  \bibinfo{journal}{Astrophys. Space Sci.} \textbf{\bibinfo{volume}{116}},
  \bibinfo{pages}{149} (\bibinfo{year}{1985}).

\bibitem[{\citenamefont{{Korycansky}}(1992)}]{korycansky92}
\bibinfo{author}{\bibfnamefont{D.~G.} \bibnamefont{{Korycansky}}},
  \bibinfo{journal}{ApJ} \textbf{\bibinfo{volume}{399}}, \bibinfo{pages}{176}
  (\bibinfo{year}{1992}).

\bibitem[{\citenamefont{{Lesur}}(2007)}]{lesurphd}
\bibinfo{author}{\bibfnamefont{G.}~\bibnamefont{{Lesur}}}, Ph.D. thesis,
  \bibinfo{school}{Universit\'e Joseph Fourier - Grenoble I}
  (\bibinfo{year}{2007}),
  \bibinfo{note}{\url{http://tel.archives-ouvertes.fr/docs/00/16/60/16/PDF/the%
se.pdf}}.

\bibitem[{\citenamefont{{Lesur} and {Longaretti}}(2005)}]{lesur05}
\bibinfo{author}{\bibfnamefont{G.}~\bibnamefont{{Lesur}}} \bibnamefont{and}
  \bibinfo{author}{\bibfnamefont{P.-Y.} \bibnamefont{{Longaretti}}},
  \bibinfo{journal}{A{\&}A} \textbf{\bibinfo{volume}{444}}, \bibinfo{pages}{25}
  (\bibinfo{year}{2005}).

\bibitem[{\citenamefont{Balay et~al.}(2011)\citenamefont{Balay, Brown,
  Buschelman, Gropp, Kaushik, Knepley, McInnes, Smith, and Zhang}}]{petsc}
\bibinfo{author}{\bibfnamefont{S.}~\bibnamefont{Balay}},
  \bibinfo{author}{\bibfnamefont{J.}~\bibnamefont{Brown}},
  \bibinfo{author}{\bibfnamefont{K.}~\bibnamefont{Buschelman}},
  \bibinfo{author}{\bibfnamefont{W.~D.} \bibnamefont{Gropp}},
  \bibinfo{author}{\bibfnamefont{D.}~\bibnamefont{Kaushik}},
  \bibinfo{author}{\bibfnamefont{M.~G.} \bibnamefont{Knepley}},
  \bibinfo{author}{\bibfnamefont{L.~C.} \bibnamefont{McInnes}},
  \bibinfo{author}{\bibfnamefont{B.~F.} \bibnamefont{Smith}}, \bibnamefont{and}
  \bibinfo{author}{\bibfnamefont{H.}~\bibnamefont{Zhang}},
  \emph{\bibinfo{title}{{PETSc} {W}eb page}} (\bibinfo{year}{2011}),
  \bibinfo{note}{\url{http://www.mcs.anl.gov/petsc}}.

\bibitem[{\citenamefont{Hernandez et~al.}(2005)\citenamefont{Hernandez, Roman,
  and Vidal}}]{slepc}
\bibinfo{author}{\bibfnamefont{V.}~\bibnamefont{Hernandez}},
  \bibinfo{author}{\bibfnamefont{J.~E.} \bibnamefont{Roman}}, \bibnamefont{and}
  \bibinfo{author}{\bibfnamefont{V.}~\bibnamefont{Vidal}},
  \bibinfo{journal}{ACM Transactions on Mathematical Software}
  \textbf{\bibinfo{volume}{31}}, \bibinfo{pages}{351} (\bibinfo{year}{2005}).

\bibitem[{\citenamefont{{Lan} and {Cvitanovi{\'c}}}(2008)}]{lan08}
\bibinfo{author}{\bibfnamefont{Y.}~\bibnamefont{{Lan}}} \bibnamefont{and}
  \bibinfo{author}{\bibfnamefont{P.}~\bibnamefont{{Cvitanovi{\'c}}}},
  \bibinfo{journal}{Phys. Rev. E} \textbf{\bibinfo{volume}{78}},
  \bibinfo{pages}{026208} (\bibinfo{year}{2008}).

\bibitem[{\citenamefont{{Lesur} and {Ogilvie}}(2008{\natexlab{b}})}]{lesur08b}
\bibinfo{author}{\bibfnamefont{G.}~\bibnamefont{{Lesur}}} \bibnamefont{and}
  \bibinfo{author}{\bibfnamefont{G.~I.} \bibnamefont{{Ogilvie}}},
  \bibinfo{journal}{MNRAS} \textbf{\bibinfo{volume}{391}},
  \bibinfo{pages}{1437} (\bibinfo{year}{2008}{\natexlab{b}}).

\bibitem[{\citenamefont{{Balbus} and {Hawley}}(1992)}]{balbus92}
\bibinfo{author}{\bibfnamefont{S.~A.} \bibnamefont{{Balbus}}} \bibnamefont{and}
  \bibinfo{author}{\bibfnamefont{J.~F.} \bibnamefont{{Hawley}}},
  \bibinfo{journal}{ApJ} \textbf{\bibinfo{volume}{400}}, \bibinfo{pages}{610}
  (\bibinfo{year}{1992}).

\bibitem[{\citenamefont{{Ogilvie} and {Pringle}}(1996)}]{ogilvie96}
\bibinfo{author}{\bibfnamefont{G.~I.} \bibnamefont{{Ogilvie}}}
  \bibnamefont{and} \bibinfo{author}{\bibfnamefont{J.~E.}
  \bibnamefont{{Pringle}}}, \bibinfo{journal}{MNRAS}
  \textbf{\bibinfo{volume}{279}}, \bibinfo{pages}{152} (\bibinfo{year}{1996}).

\bibitem[{\citenamefont{{Terquem} and {Papaloizou}}(1996)}]{terquem96}
\bibinfo{author}{\bibfnamefont{C.}~\bibnamefont{{Terquem}}} \bibnamefont{and}
  \bibinfo{author}{\bibfnamefont{J.~C.~B.} \bibnamefont{{Papaloizou}}},
  \bibinfo{journal}{MNRAS} \textbf{\bibinfo{volume}{279}}, \bibinfo{pages}{767}
  (\bibinfo{year}{1996}).

\bibitem[{\citenamefont{{Butler} and {Farrell}}(1992)}]{butler92}
\bibinfo{author}{\bibfnamefont{K.~M.} \bibnamefont{{Butler}}} \bibnamefont{and}
  \bibinfo{author}{\bibfnamefont{B.~F.} \bibnamefont{{Farrell}}},
  \bibinfo{journal}{Phys. Fluids} \textbf{\bibinfo{volume}{4}},
  \bibinfo{pages}{1637} (\bibinfo{year}{1992}).

\bibitem[{\citenamefont{{Johnson}}(2007)}]{johnson07}
\bibinfo{author}{\bibfnamefont{B.~M.} \bibnamefont{{Johnson}}},
  \bibinfo{journal}{ApJ} \textbf{\bibinfo{volume}{660}}, \bibinfo{pages}{1375}
  (\bibinfo{year}{2007}).

\bibitem[{\citenamefont{{Farrell} and {Ioannou}}(1993)}]{farrell93}
\bibinfo{author}{\bibfnamefont{B.~F.} \bibnamefont{{Farrell}}}
  \bibnamefont{and} \bibinfo{author}{\bibfnamefont{P.~J.}
  \bibnamefont{{Ioannou}}}, \bibinfo{journal}{Phys. Fluids}
  \textbf{\bibinfo{volume}{5}}, \bibinfo{pages}{1390} (\bibinfo{year}{1993}).

\bibitem[{\citenamefont{{Lithwick}}(2007)}]{lithwick07}
\bibinfo{author}{\bibfnamefont{Y.}~\bibnamefont{{Lithwick}}},
  \bibinfo{journal}{ApJ} \textbf{\bibinfo{volume}{670}}, \bibinfo{pages}{789}
  (\bibinfo{year}{2007}).

\bibitem[{\citenamefont{{Nagata}}(1986)}]{nagata86}
\bibinfo{author}{\bibfnamefont{M.}~\bibnamefont{{Nagata}}},
  \bibinfo{journal}{J. Fluid Mech.} \textbf{\bibinfo{volume}{169}},
  \bibinfo{pages}{229} (\bibinfo{year}{1986}).

\bibitem[{\citenamefont{{Gressel}}(2010)}]{gressel10}
\bibinfo{author}{\bibfnamefont{O.}~\bibnamefont{{Gressel}}},
  \bibinfo{journal}{MNRAS} \textbf{\bibinfo{volume}{405}}, \bibinfo{pages}{41}
  (\bibinfo{year}{2010}).

\bibitem[{\citenamefont{{Davies} and {Hughes}}(2011)}]{davies10}
\bibinfo{author}{\bibfnamefont{C.~R.} \bibnamefont{{Davies}}} \bibnamefont{and}
  \bibinfo{author}{\bibfnamefont{D.~W.} \bibnamefont{{Hughes}}},
  \bibinfo{journal}{ApJ} \textbf{\bibinfo{volume}{727}}, \bibinfo{pages}{112}
  (\bibinfo{year}{2011}).

\bibitem[{\citenamefont{{Miesch} et~al.}(2007)\citenamefont{{Miesch}, {Gilman},
  and {Dikpati}}}]{miesch07}
\bibinfo{author}{\bibfnamefont{M.~S.} \bibnamefont{{Miesch}}},
  \bibinfo{author}{\bibfnamefont{P.~A.} \bibnamefont{{Gilman}}},
  \bibnamefont{and}
  \bibinfo{author}{\bibfnamefont{M.}~\bibnamefont{{Dikpati}}},
  \bibinfo{journal}{ApJ. Supp. Ser.} \textbf{\bibinfo{volume}{168}},
  \bibinfo{pages}{337} (\bibinfo{year}{2007}).

\bibitem[{\citenamefont{{Cvitanovic}}(1992)}]{cvit92}
\bibinfo{author}{\bibfnamefont{P.}~\bibnamefont{{Cvitanovic}}},
  \bibinfo{journal}{Chaos} \textbf{\bibinfo{volume}{2}}, \bibinfo{pages}{1}
  (\bibinfo{year}{1992}).

\end{thebibliography}

\onecolumngrid
\appendix*
\section{Symmetries\label{appsym}}
Nagata \citep{nagata86} identified several possible symmetries for
three-dimensional nonlinear hydrodynamic solutions in centrifugally
unstable (Rayleigh-unstable) Taylor-Couette flow in the thin-gap limit,
which corresponds to a cartesian plane Couette flow with walls,
rotating along its spanwise $z$ axis. These symmetries can be 
adapted to the shearing box and generalized to MHD flows. 

We consider a physical domain with $0\leq x<L_x$, $0\leq y<L_y$ and $0\leq z<L_z$. 
For visualization purposes, we assume that the observer is comoving 
with the base flow at $x=L_x/2$ and therefore introduce 
\begin{equation}
  \label{eq:xprime}
  y'= y-\f{L_x}{2}  St~.
\end{equation}
In Fig.~\ref{figure4}, the coordinates of the lower left corner of each box 
are $x=0$, $y'=0$, $z=0$. The various fields entering the nonlinear dynamo cycle 
solution presented in this paper have the following generalized $\mathcal{A}_1$
symmetry:
\newcommand{\rme}{\mathrm{e}}
\newcommand{\rmo}{\mathrm{o}}
\begin{equation}
\vec{u}:\left\{
\begin{array}{lllll}
u_x & = & 
u_{x,\rme\rme}(t)\cos\left(k_{z\rme}z\right)\sin\left(k_{y\rme}y'+k_x(t)x\right)
& +&
u_{x,\rmo\rmo}(t)\cos\left(k_{z\rmo}z\right)\cos\left(k_{y\rmo}y'+k_x(t)x\right)\\
u_y & = & 
u_{y,\rme\rme}(t)\cos\left(k_{z\rme}z\right)\sin\left(k_{y\rme}y'+k_x(t)x\right)
& +& 
u_{y,\rmo\rmo}(t)\cos\left(k_{z\rmo}z\right)\cos\left(k_{y\rmo}y'+k_x(t)x\right)\\
u_z  & = & 
u_{z,\rme\rme}(t)\sin\left(k_{z\rme}z\right)\cos\left(k_{y\rme}y'+k_x(t)x\right)
& + &
u_{z,\rmo\rmo}(t)\sin\left(k_{z\rmo}z\right)\sin\left(k_{y\rmo}y'+k_x(t)x\right)
\end{array}\right\}~,
\end{equation}
\begin{equation}
\vec{B}:\left\{
\begin{array}{lllll}
B_x & = & 
B_{x,\rme\rmo}(t)\cos\left(k_{z\rme}z\right)\sin\left(k_{y\rmo}y'+k_x(t)x\right) 
& + &
B_{x,\rmo\rme}(t)\cos\left(k_{z\rmo}z\right)\cos\left(k_{y\rme}y'+k_x(t)x\right) \\
B_y & = & 
B_{y,\rme\rmo}(t)\cos\left(k_{z\rme}z\right)\sin\left(k_{y\rmo}y'+k_x(t)x\right) 
& + &
B_{y,\rmo\rme}(t)\cos\left(k_{z\rmo}z\right)\cos\left(k_{y\rme}y'+k_x(t)x\right) \\
B_z & = &
B_{z,\rme\rmo}(t)\sin\left(k_{z\rme}z\right)\cos\left(k_{y\rmo}y'+k_x(t)x\right)
& + &
B_{z,\rmo\rme}(t)\sin\left(k_{z\rmo}z\right)\sin\left(k_{y\rme}y'+k_x(t)x\right)
\end{array}\right\}~,
\end{equation}
where the $\rme$ and $\rmo$ subscripts indicate that the associated discrete 
wavenumbers are based on even and odd relative integers, respectively. $k_x(t)$ 
in each individual trigonometric expression is defined implicitly by 
Eq.~(\ref{eq:shwave}) using the $k_y$ wavenumber of the same expression. 
It can be checked that this symmetry is conserved by the MHD
equations~(\ref{eq:NS})-(\ref{eq:div}). When needed, we enforced
numerically that the dynamical evolution took place in the invariant
subspace defined by this symmetry by imposing it in the initial
conditions and by further enforcing it every shearing time during the
numerical integrations in order to avoid the growth 
of non-symmetric numerical noise. 

\end{document}